# MosaicJoin: Compact Semantic Sketches for Value-Level Join Discovery

Grace Fan
New York University
grace.fan@nyu.edu

Eden Wu
New York University
eden.wu@nyu.edu

Majid Daliri
New York University
daliri.majid@nyu.edu

Juliana Freire
New York University
juliana.freire@nyu.edu

## ABSTRACT

Join discovery is a core task in dataset search, enabling users to find columns that can be joined with a given query column. Early approaches focused on equi-joins, but data lakes and open-data repositories often contain columns whose values refer to the same entity but use different syntactic representations. To address this challenge, recent approaches discover semantically joinable columns but face a fundamental trade-off: methods that perform value-level comparisons accurately identify joinable columns but scale poorly to columns with high cardinality; column-level methods that encode an entire column into a single embedding are efficient but do not capture the fine-grained value alignment that determines whether a join is possible. We present MosaicJoin, a value-level semantic join discovery method that balances this trade-off. MosaicJoin achieves scalability through a novel sketching strategy that approximates the joinability of a column pair without having to compare all values. At query time, MosaicJoin scores each candidate sketch using a joinability score at a cost bounded by the sketch size, making retrieval efficient even for high-cardinality columns. A query subsampling operator further reduces online search time with provable accuracy guarantees, enabling robust retrieval for large query columns. Extensive experiments show that MosaicJoin outperforms previously published methods across all benchmarks while running up to 66 times faster than other value-level methods. MosaicJoin requires no training or fine-tuning, and it scales robustly to query columns containing up to 57K values and data lake columns containing up to 1M values.

**PVLDB Reference Format:**
Grace Fan, Eden Wu, Majid Daliri, and Juliana Freire. MosaicJoin: Compact Semantic Sketches for
Value-Level Join Discovery. PVLDB, 19(11): XXX-XXX, 2026.
doi:XX.XX/XXX.XX

**PVLDB Artifact Availability:**
The source code, data, and/or other artifacts have been made available at https://github.com/VIDA-NYU/MosaicJoin.

## 1 INTRODUCTION

The number of datasets in open and enterprise data repositories and data lakes has grown dramatically in recent years [47, 53], creating opportunities to augment data, improve machine learning models, and enrich analyses [7, 52]. Yet the scale and heterogeneity of these repositories make manual data discovery infeasible, as metadata in data lakes is frequently incomplete, inconsistent, or ambiguous [1, 25, 57, 72]. Automated data discovery has thus attracted significant attention [8, 22, 29, 57], with join discovery [64, 67, 74, 76] emerging as a key subtask. In particular, finding *semantically* joinable columns [12, 18, 19, 31, 43] is essential for data enrichment [6, 9, 17, 73].

Given a query column and a large repository, join discovery aims to identify columns whose values can be joined with those in the query column. Early systems focused on equi-joins, finding columns that share exact string values with the query [27, 64, 74, 76]. However, these approaches suffer from low recall in real data lakes, where semantically equivalent values frequently differ in surface form through abbreviations, aliases, and misspellings.

*Semantic Joins.* To handle such variations, existing semantic join discovery methods determine joinability based on either column-level or value-level similarities. Value-level methods such as PEXESO [18] and Exact Semantic Join, an exhaustive value-embedding baseline we evaluate in Section 5, compare individual value embeddings across columns, directly measuring fine-grained joinability. However, exhaustive value-pair comparison scales poorly: for columns with tens of thousands of values, the cost becomes prohibitive, as shown in Figure 1. Column-level methods such as DeepJoin [19] and Snoopy [31], by contrast, collapse an entire column into a single fixed-length embedding retrieved via approximate nearest neighbor search. While efficient, a single vector cannot capture the fine-grained value alignment that determines whether a join is possible, producing both false negatives — joinable columns whose values align but whose column-level embeddings do not, and false positives — columns that are topically similar but not value-level joinable.

Example 1. *Suppose a user has a query column about football league seasons ($C_Q$, top left in Figure 2). Equi-join methods miss the green column $C_G$ because its values are syntactically different: "2003 Tippeligaen" is the Norwegian name for 2003 Norwegian Premier League, "1992 Vyshcha Liha" is the predecessor name of 1992 Ukrainian Premier League, and "2006 Division 1 (Swedish football)" is a formatting variant of 2006 Swedish football Division 1. Column-level methods instead return the blue column $C_B$ that contains basketball team names, which is topically related to sports but whose values do not align with football league seasons.*

*Our Approach.* We propose MosaicJoin, a value-driven semantic join discovery method that achieves high accuracy of value-level matching while approaching the efficiency of column-level retrieval, as shown in Figure 1. Compared to the exhaustive value-pair comparison method for joinability (Exact Semantic Join), MosaicJoin reduces online search time from 15.65 seconds to 0.32 seconds while preserving high accuracy. This sub-second response time is important both for interactive dataset search [7, 32], where users explore join candidates in real time, and for automated data integration pipelines [6, 9], where a large number of potential join pairs must be evaluated without the discovery process becoming a bottleneck. The key insight behind MosaicJoin is that determining the joinability of two columns does not require comparing all pairs of value embeddings. Instead, for each data lake column, MosaicJoin constructs a fixed-size set of representative value embeddings–*a semantic sketch*–that maximally covers the column's embedding space via $k$-center sampling [36]. By selecting representatives that minimize the maximum distance from any value embedding to its nearest sketch point, the k-center objective guarantees that the full embedding space of the column is covered, so that similarity computed against the sketch closely approximates similarity computed against all values. At query time, MosaicJoin scores each candidate sketch using Chamfer-style similarity score [36] at a cost bounded by sketch size rather than data lake column cardinality, making retrieval efficient even for columns with a very large number of values. To further reduce online query cost, rather than embedding all query values, we subsample the values and use the resulting subset to estimate the full Chamfer score with provable accuracy guarantees. Unlike deep learning-based approaches [12, 19, 31, 34, 38, 43, 49], MosaicJoin requires no training or fine-tuning and can be deployed immediately on any new or evolving data lake without task-specific retraining.

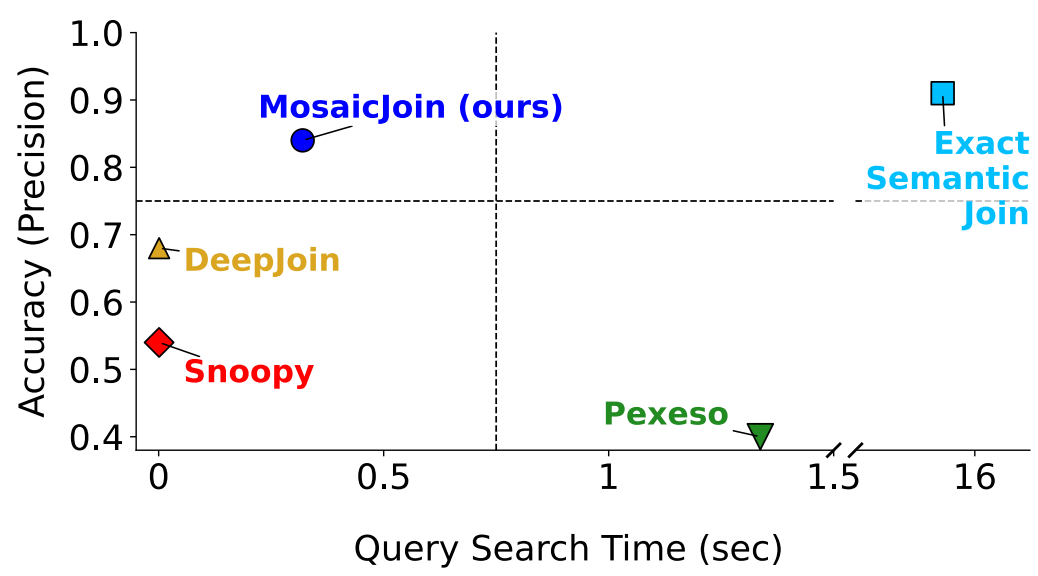

**Figure 1: Tradeoff between Accuracy of Value-level Joinability and Online Search Efficiency of Column-level joinability, for query columns with up to 57K values (Section 5).**

We make the following contributions:

- We propose MosaicJoin, a training-free semantic join discovery method that preserves value-level joinability evidence while enabling efficient online top-$k$ retrieval over data lakes.
- We introduce compact semantic sketches for columns, constructed with a $k$-center procedure that covers the column embedding space and preserves rare but join-critical values.
- We define a Chamfer-style joinability score over semantic sketches, bounding candidate comparison cost by sketch size rather than column cardinality. We further reduce query-time cost using a subsampling estimator with provable accuracy guarantees.
- We introduce an evaluation protocol for semantic join discovery on fuzzy-join benchmarks and WDC-augmented search spaces, using LLM-as-a-judge silver labels with human validation to assess value-level alignment at scale [70].
- We evaluate MosaicJoin on six benchmarks with up to 99K columns and 1M values and show that it outperforms previously published methods by up to 17.6% in the ranking metric NDCG@20 and 135.7% in the accuracy metric Precision@10, while being up to 66× faster than value-level baselines.

## 2 RELATED WORK

**Table Discovery.** Table discovery has a rich line of literature [8, 22, 29, 57]. Early work focused on keyword search over the metadata of tables [4, 5, 32], while recent systems support interactive exploration via data profiling and relationship discovery [7, 21, 26, 53, 55, 56, 61, 72]. Relationships include joinability, unionability, and correlation to improve machine learning models and explain data [2, 3, 10, 13, 23, 35, 37, 58, 64, 67]. We focus on join discovery. Equi-join approaches approximate Jaccard similarity and set containment to find value overlap [27, 74, 76], with later techniques supporting joinable and correlated tables [64–66] and multi-attribute joins [20]. These only handle exact string overlap, missing semantically equivalent joins. More recent work has incorporated semantics into join discovery [16, 40]. PEXESO [18] performs exact value-level semantic join discovery by embedding cells individually and retrieving candidates within a distance threshold. Although PEXESO utilizes pivot-based pruning and a block-and-verify index, it remains verification-heavy with runtime scaling in both query length and the number of filtered candidates; we include PEXESO as an exact value-level baseline in our experiments. Column-level join methods such as DeepJoin [19] and WarpGate [12] encode each column into a single fixed-length embedding via a pretrained language model and retrieve candidates via approximate nearest-neighbor (ANN) search. While efficient, they must truncate high-cardinality columns, which can fail to capture value-level joinability; we use DeepJoin as the representative column-level baseline; we use DeepJoin as the representative column-level baseline. Snoopy [31] performs efficient value-level joinability by projecting a column onto learned proxy columns to produce a fixed-dimensional embedding for ANN retrieval. Since Snoopy's objective is most closely aligned with ours, we compare MosaicJoin against Snoopy directly.

Several related efforts address distinct settings. SILKMOTH [15] and KOIOS [54] perform semantic overlap search via maximum bipartite matching using filter-verification frameworks. While related to join discovery, SILKMOTH and KOIOS require exact maximum bipartite matching at query time, making it impractical for noisy data lakes that do not contain only one-to-one joins. Nonetheless, we include SILKMOTH and KOIOS in our experiments. OmniMatch [43] and HyperJoin [49] both create similarity graphs over all columns, discovering all joins fully offline rather than performing online top-$k$ retrieval as MosaicJoin does. Freyja [50] constructs column-level profiles that do not capture fine-grained value-level alignment, and PolyJoin [34] targets multi-key (n-ary) join search. TabSketchFM [38] uses sketch-based table representations capturing only exact value overlap and numerical similarity, whereas MosaicJoin captures the semantic alignment of values in embedding space.[1] TOPJoin [41] targets context-aware joinability in enterprise settings where valid joins require table-level semantic relationships, not just value alignments; we include it as a context-aware baseline. Broadly, these approaches either require offline, pairwise-heavy modeling or compress columns into coarse heuristic profiles, whereas MosaicJoin uses bounded, deterministic value sketches to keep value-level semantic join evidence while supporting fast online top-$k$ retrieval, even for high-cardinality columns.

[1]The model is not publicly available for TabSketchFM, so we are not able to include it in our experiments.

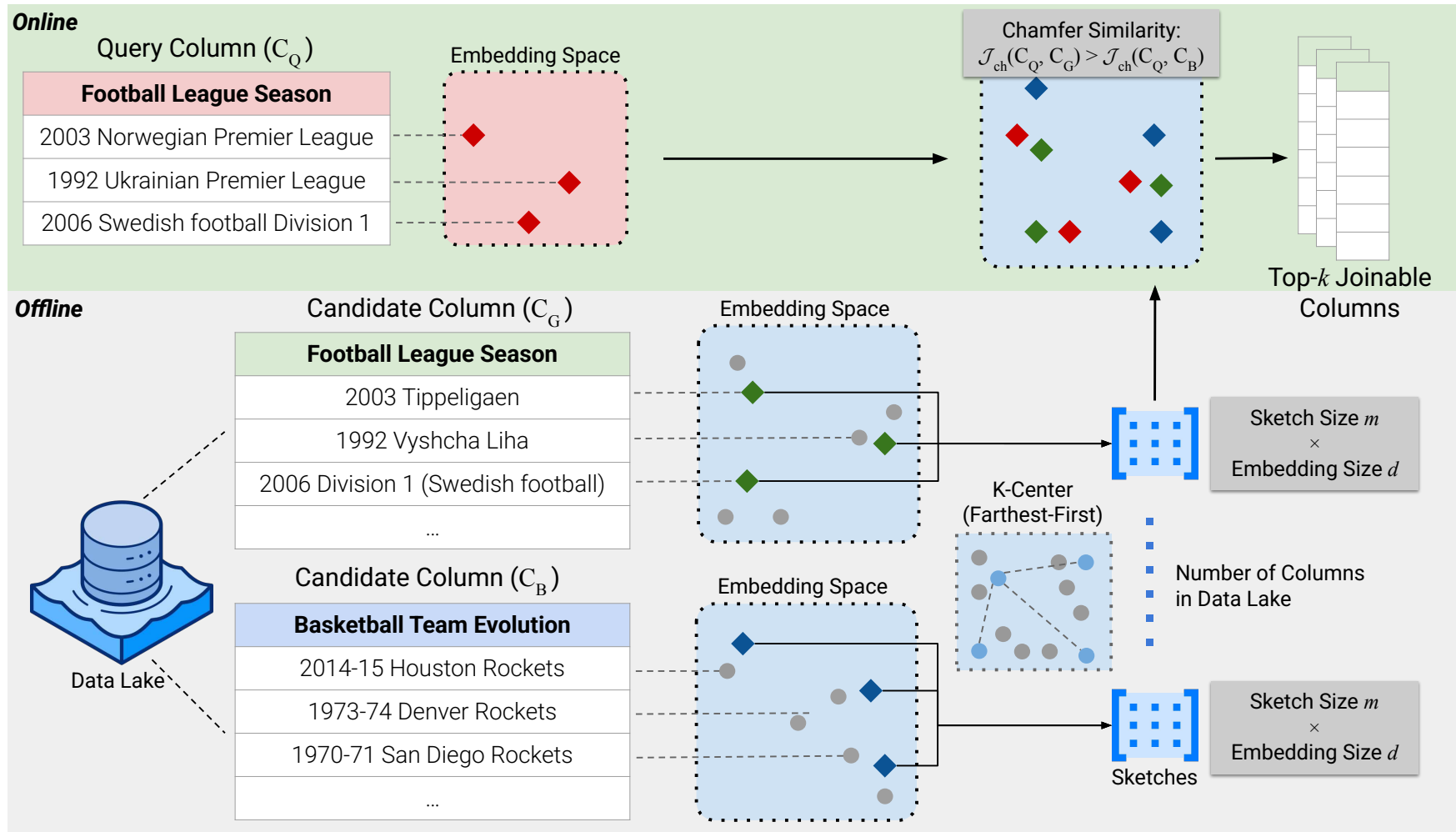

**Figure 2: MosaicJoin Pipeline, with real columns from our experimental evaluations [48] (Section 5). During the offline phase, MosaicJoin embeds column values and creates a compact semantic sketch using the $k$-center procedure. During online processing, given a query column, MosaicJoin embeds the column values and ranks candidate columns using a Chamfer-style semantic joinability score. MosaicJoin correctly returns the green column ($C_G$) about Football league seasons, whose values match those in the red query column ($C_Q$) in the top left. Column-based methods return the bottom blue column ($C_B$), whose column semantics are similar to those of the query column (basketball and football), but whose values do not align.**

**Fuzzy-Joins.** A complementary line of work studies approximate (fuzzy) joins between two given tables by automatically selecting similarity functions and transformation rules [59]. Auto-FuzzyJoin auto-programs a fuzzy-join to meet a user-specified precision target while maximizing recall, alleviating manual tuning of distance functions and thresholds [48]. Auto-Join [75] learns string transformation programs so that transformed keys can be equi-joined. More recently, DTT [14] uses neural language modeling to map entity columns into joinable formats, avoiding exhaustive searches over hand-crafted transformation spaces. These techniques focus on pairwise join execution or key transformation between two given tables, whereas MosaicJoin searches a large repository for columns whose values can be joined with a query column.

**Multivector Search.** Late-interaction retrievers represent queries and documents as *sets* of vectors and score relevance by aggregating fine-grained per-element matches, enabling token-level accuracy while allowing document representations to be precomputed for efficiency. ColBERT [39] popularized this design, demonstrating that it preserves token-level evidence and is more robust than collapsing a whole text into a single embedding. However, exact evaluation is expensive because it requires many cross-vector comparisons and cannot be reduced to a single maximum inner product search (MIPS) query. MUVERA [36] addresses this by constructing fixed-dimensional encodings whose dot product approximates the underlying set-based similarity, enabling candidate retrieval via off-the-shelf MIPS indices with lightweight re-ranking. These results motivate our use of aggregate-of-best-matches scoring: it preserves fine-grained evidence about which query values are supported by a candidate column while admitting efficient top-k retrieval through through indexing and approximation mechanisms.

## 3 PRELIMINARIES

In this section, we introduce our notion of semantic joinability and the compact column representation used to make retrieval efficient at data-lake scale. We first define a Chamfer-style semantic joinability score for ranking candidate columns, and then motivate semantic sketches constructed via a $k$-center objective.

### 3.1 Semantic Joinability

Traditional join discovery is often based on *exact value overlap*: two columns are considered joinable if many values in one column appear verbatim in the other [27, 64, 74, 76]. While effective for clean, normalized data, this criterion is brittle in data lakes, where semantically equivalent values may differ in surface through abbreviations, aliases, formatting differences, misspellings.

To address this, we move from *exact matching* to *semantic matching*. Instead of comparing raw strings directly, we compare value embeddings produced by an embedding function $\phi(\cdot)$, so that values with similar meaning are close in $\mathbb{R}^d$ even when their string forms differ. This allows us to determine whether a value in the query column has a semantically similar counterpart in a candidate column.

Let $\Phi(C)$ denote the multiset of value embeddings for column $C$:

$$\Phi(C) = \{\phi(v) \mid v \in C\}$$

*String-Based Exact-Match Joinability.* A strict exact-match score can be written as

$$\mathcal{J}_{\text{exact}}(C_Q, C) = \frac{1}{|C_Q|} \sum_{v \in C_Q} \mathbf{1}[\exists v' \in C \text{ such that } v = v'] \quad (1)$$

This measures the fraction of query values that appear exactly in $C$.
*Semantic Joinability via a Directed Chamfer-Style Score.* Given a similarity function sim (e.g., cosine similarity), we define semantic joinability as

$$\mathcal{J}_{\text{ch}}(C_Q, C) = \frac{1}{|C_Q|} \sum_{v \in C_Q} \max_{u \in \Phi(C)} \text{sim}(\phi(v), u). \quad (2)$$

This is a directed Chamfer-style aggregation [36]: for each value in the query column $C_Q$, we keep only the best semantic match in the candidate column – the embedding $u$ with the highest similarity – and then, compute the average across all values.

This formulation reflects the semantics of joins in real data lakes, where value mappings are frequently non-bijective. For example, in a city-to-state join, several city names may map to the same state. A one-to-one bipartite matching objective would allow only one city to claim a state, discarding the remaining join evidence and underestimating joinability. The directed Chamfer score avoids this by letting each query value independently contribute its best-match similarity, so $\mathcal{J}_{\text{ch}}$ measures semantic coverage of the query by the candidate column. This many-to-one modeling is not an assumption unique to MosaicJoin, it is independently motivated by SEMA-JOIN [33], which studies semantic joins beyond exact equality and explicitly models optional many-to-one mappings as a core join pattern. The distinction is one of setting: SEMA-JOIN predicts an instance-level mapping between two given columns using co-occurrence statistics from a large table corpus, whereas MosaicJoin targets top-k joinable-column retrieval over a data lake, where scores must be efficiently computed across many candidates and require no corpus-level supervision. The directed Chamfer score satisfies both requirements simultaneously.
*Ranking candidate columns.* Given a query column $C_Q$ and a set of candidate columns $\mathcal{C}$, the score $\mathcal{J}_{\text{ch}}(C_Q, C)$ provides a common scale for comparison. In particular, if

$$\mathcal{J}_{\text{ch}}(C_Q, C_a) > \mathcal{J}_{\text{ch}}(C_Q, C_b),$$

then, on average, values in $C_Q$ have stronger best semantic matches in $C_a$ than in $C_b$. Equivalently, $C_a$ provides better semantic coverage of the query values than $C_b$.

Therefore, top-$k$ semantic joinable column search can be implemented by ranking candidate columns by $\mathcal{J}_{\text{ch}}(C_Q, C)$ and returning the highest-scoring ones. Columns with high scores provide strong semantic coverage of the query, i.e., most query values have at least one close semantic match and are therefore strong join candidates regardless of whether the value mapping is one-to-one or one-to-many.

### 3.2 Compact Semantic Sketches via $k$-Center

*The Need for Compact Representations.* Although $\mathcal{J}_{\text{ch}}$ is a natural semantic analogue of exact overlap, computing it exactly is prohibitively expensive at data-lake scale. For a data lake with N candidate columns, evaluating $\mathcal{J}_{\text{ch}}$ exactly requires computing, for each query value $v \in C_Q$, the maximum similarity against every value embedding in every candidate column. This yields an online cost of $O(Nn|C|d)$, where $n = |C_Q|$ is the query column size, |C| is the candidate column size, and d is the embedding dimension — scaling with both query and data lake column cardinalities simultaneously. For the largest columns in our evaluation (up to 57K query values and 1M data lake values [50]), this cost is reflected in the 15.65-second per-query latency of ESJ, our exact baseline. To make retrieval efficient, we replace the full embedding multiset $\Phi(C)$ with a compact representative subset, which we call a *semantic sketch.*
*Semantic Sketch and Representative Embeddings.* For each column $C$ and sketch size $m$, we construct a sketch

$$S(C) \subseteq \Phi(C), \qquad |S(C)| = m \ll |C|,$$

The goal is for $S(C)$ to preserve the semantic coverage of $\Phi(C)$, so that joinability computed against the sketch remains a good approximation of joinability computed against the full column.
*$k$-center Objective for Sketch Construction.* We construct $S(C)$ using a $k$-center-style objective (with budget $m$). Let $D(\cdot, \cdot)$ be a distance corresponding to the similarity function (e.g., cosine distance when sim is cosine similarity). We choose $m$ representatives that minimize the worst-case distance from any embedding in $\Phi(C)$ to its nearest representative:

$$S(C) \in \arg \min_{\substack{S \subseteq \Phi(C) \\ |S|=m}} \max_{x \in \Phi(C)} \min_{s \in S} D(x, s). \quad (3)$$

Intuitively, this objective ensures that every value embedding in the column is close to at least one selected representative. This is well aligned with the Chamfer-style joinability score, which is based on best matches: if each original embedding is well covered by a nearby representative, then nearest-match similarities against the full column can be approximated using only the sketch.
*Approximation.* The objective above is the $k$-center problem with budget $m$, which is NP-hard in general and closely related to facility-location-style clustering. MosaicJoin therefore uses the standard greedy farthest-first traversal, described in Section 4.2, rather than solving the objective exactly. This approximation is well suited to our setting because it greedily reduces the maximum distance from any column value to its nearest representative, preserving the semantic coverage needed by the Chamfer-style best-match score. For metric distances, farthest-first gives a constant-factor approximation to the optimal $k$-center radius. This sketch is built once offline during preprocessing and reused during online scoring, so its cost does not affect query-time search efficiency.

## 4 MOSAICJOIN FRAMEWORK

We now discuss the MosaicJoin framework, including its offline stage where we build compact semantic sketches (Section 4.2) and online stage where we determine semantic joinability of a query column (Section 4.3).

### 4.1 System Overview

Figure 2 shows the overall pipeline of MosaicJoin, which consists of an offline and online stage. Given a data lake with columns $\mathcal{C}$, in the *offline stage*, we embed values in each data-lake column and compress each column into a fixed-size *semantic sketch* $S(C)$ using a k-center (farthest-first) procedure. In the *online stage*, given a query column $C_Q$, we embed its values in the same way and rank

**Algorithm 1:** COMPACT COLUMN REPRESENTATION via Farthest-First Selection

```
Input: Value embeddings Φ = {x_1, ..., x_n} ⊂ R^d, sketch
       size m
Output: Semantic sketch S ⊆ Φ with |S| = m
1 μ ← (1/n) Σ_{i=1}^n x_i     // centroid s_1 ← arg min_{x_i∈Φ} D(x_i, μ)
  // most central point S ← {s_1};
2 foreach x_i ∈ Φ do
3   δ_i ← D(x_i, s_1)   // distance to nearest selected
      point
4 for t = 2 to m do
5   s_t ← arg max_{x_i∈Φ\S} δ_i;
6   S ← S ∪ {s_t};
7   foreach x_i ∈ Φ \ S do
8     δ_i ← min(δ_i, D(x_i, s_t));
9 return S;
```

candidate columns by a sketch-based semantic joinability score. Throughout, we reserve $k$ for *top-k retrieval* and use $m$ for the *sketch size.*

## 4.2 Offline Stage

*Value Embeddings.* Let $C$ be a column with values $\{v_1, \dots, v_n\}$. To capture the semantics of each value, we form the text $t_i := [v_i]$ and compute an embedding $x_i := \phi(t_i) \in \mathbb{R}^d$ where $\phi(\cdot)$ denotes the embedding function. We denote the resulting (multi)set of value embeddings by $\Phi(C) := \{x_1, \dots, x_n\}$. Unless stated otherwise, we $\ell_2$-normalize embeddings so that $\|x_i\|_2 = 1$ and cosine similarity equals the dot product: $\text{sim}(x, y) = x^\top y \in [-1, 1]$.

*k-center Semantic Sketch Construction.* Storing all embeddings $\Phi(C)$ for every column is too expensive at a data-lake scale. Instead, we preprocess each column to derive a compact *semantic sketch*

$$S(C) \subseteq \Phi(C), \qquad |S(C)| = m \ll |\Phi(C)|.$$

Intuitively, $S(C)$ should preserve the semantic *coverage* of $\Phi(C)$, so that online scoring can compare a query column against sketches rather than all values.

To obtain diverse representatives, we use a k-center style farthest-first procedure [36]. Let $D(\cdot, \cdot)$ be a distance consistent with our similarity; with normalized embeddings we use cosine distance $D(x, y) := 1 - x^\top y$. The k-center objective seeks a subset $S$ of the embeddings with size $m$ that minimizes the worst-case distance from any point to its nearest representative:

$$\min_{S \subseteq \Phi(C),\, |S|=m} \; \max_{x \in \Phi(C)} \; \min_{s \in S} D(x, s).$$

We approximate this objective with farthest-first traversal, as described in Algorithm 1: we seed the sketch with the most central embedding (closest to the centroid), then repeatedly add the point farthest from its nearest selected representative. This encourages broad coverage of the embedding space and empirically preserves rare-but-informative values.

*Implementation Note.* Algorithm 1 runs in $O(nm)$ distance computations per column using the maintained nearest-center distances $\{\delta_i\}$, and is easily parallelized across columns in the offline pipeline.

**Algorithm 2:** TOP-$k$ SEMANTIC JOINABLE COLUMN RETRIEVAL

```
Input: Query column C_Q; candidate columns C; embedding
       function φ(·); compact reps {R(C)}; return size k;
       query sample size b
Output: Top-k columns ranked by J~_ch^(b)(C_Q, C)
1  Q ← {φ(v) | v ∈ C_Q};
2  Sample a multiset Q_b ⊆ Q of size b uniformly at random
     (with replacement);
3  Initialize an empty min-heap H (capacity k) storing
     (score, C);
4  foreach C ∈ C do
5    s ← 0;
6    foreach q ∈ Q_b do
7      s ← s + max_{r∈R(C)} ⟨q, r⟩;
8    s ← s/b        // query-subsampled Chamfer score
9    if |H| < k then
10     HeapPush(H, (s, C));
11   else if s > min(H) then
12     HeapPopMin(H);
13     HeapPush(H, (s, C));
14 return HeapItemsSortedDesc(H);
```

## 4.3 Online Query Processing

Algorithm 2 specifies how queries are evaluated. Given a query column $C_Q$, we embed its values using the same embedding function $\phi(\cdot)$ and form the query embedding multiset $\Phi(C_Q)$. We then compare $\Phi(C_Q)$ against the semantic sketch $S(C)$ of each candidate column $C$ and rank candidates by their sketch-based semantic joinability score. We return the top-$k$ highest-scoring columns, optionally filtering out candidates whose score is below a fixed threshold $\tau$ (we use $\tau = 0.1$).

*Query Subsampling for Faster Scoring.* To further reduce online cost, we estimate the directed sketch-based Chamfer score using a random subset of query values. Let $C_Q = \{v_1, \dots, v_n\}$, where $n = |C_Q|$, and $S(C)$ be a candidate sketch derived using Algorithm 1.

$$a_i := \max_{s \in S(C)} \big(\phi(v_i) \cdot s\big), \qquad i = 1, \dots, n.$$

Then the full sketch-based directed Chamfer score is

$$\widehat{\mathcal{J}}_{\text{ch}}(C_Q, C) = \frac{1}{n} \sum_{i=1}^{n} a_i.$$

We sample $b$ indices $I_1, \dots, I_b$ independently and uniformly from $C_Q$ values $\{1, \dots, n\}$, and define the subsampled estimator

$$\widetilde{\mathcal{J}}_{\text{ch}}^{(b)}(C_Q, C) = \frac{1}{b} \sum_{t=1}^{b} a_{I_t}.$$

THEOREM 2 (99.9% CONFIDENCE FOR QUERY-SUBSAMPLED CHAMFER). *Fix a candidate sketch $S(C)$, assume embeddings are normalized so that $\phi(v_i) \cdot s \in [-1, 1]$ for all $i$ and $s \in S(C)$. Then*

$$\mathbb{E}\Big[\widetilde{\mathcal{J}}_{\text{ch}}^{(b)}(C_Q, C)\Big] = \widehat{\mathcal{J}}_{\text{ch}}(C_Q, C),$$

*and with probability at least* 99.9%*,*

$$\left|\widetilde{\mathcal{J}}_{\text{ch}}^{(b)}(C_Q, C) - \widehat{\mathcal{J}}_{\text{ch}}(C_Q, C)\right| = O\left(\sqrt{\frac{1}{b}}\right).$$

Proof. Define $X_t := a_{I_t}$ for $t = 1, \ldots, b$. Then

$$\widetilde{\mathcal{J}}_{\text{ch}}^{(b)}(C_Q, C) = \frac{1}{b}\sum_{t=1}^{b} X_t.$$

Since each $I_t$ is uniform on $\{1, \ldots, n\}$,

$$\mathbb{E}[X_t] = \frac{1}{n}\sum_{i=1}^{n} a_i = \widehat{\mathcal{J}}_{\text{ch}}(C_Q, C).$$

Hence, by linearity of expectation,

$$\mathbb{E}\left[\widetilde{\mathcal{J}}_{\text{ch}}^{(b)}(C_Q, C)\right] = \widehat{\mathcal{J}}_{\text{ch}}(C_Q, C).$$

Because embeddings are normalized, each $a_i \in [-1, 1]$, and therefore $X_t \in [-1, 1]$. The variables $X_1, \ldots, X_b$ are independent since the indices $I_1, \ldots, I_b$ are sampled independently. By Hoeffding's inequality, for any $\varepsilon > 0$,

$$\Pr\left(\left|\widetilde{\mathcal{J}}_{\text{ch}}^{(b)}(C_Q, C) - \widehat{\mathcal{J}}_{\text{ch}}(C_Q, C)\right| \geq \varepsilon\right) \leq 2\exp\left(-\frac{b\varepsilon^2}{2}\right).$$

Setting the failure probability to 0.001 (i.e., confidence 99.9%) gives $2\exp\left(-\frac{b\varepsilon^2}{2}\right) \leq 0.001$, which implies $\varepsilon \geq \sqrt{\frac{2\ln(2000)}{b}}$. Since $2\ln(2000)$ is a constant, the error bound at confidence 99.9% is $O\left(\sqrt{\frac{1}{b}}\right)$. □

*Query Search Complexity.* Let $N := |C|$ be the number of candidate columns, $n := |C_Q|$ be the number of query values, and assume each candidate sketch has size at most $m$ (i.e., $|S(C)| \leq m$). Given a subsample of size $b$, scoring a single candidate column requires $O(bmd)$ time, since for each of the $b$ sampled query embeddings we compute the maximum inner product against its $m$ sketch vectors in $\mathbb{R}^d$. Maintaining the top-$k$ heap adds $O(\log k)$ time per candidate. Therefore, the total online query time is

$$O(N\, bmd + N\log k).$$

When $k \ll N$, the dominant term is typically $O(N\, bmd)$, and without query subsampling (i.e., $b = n$) this becomes $O(N\, nmd)$.

# 5 EXPERIMENTS

We evaluate MosaicJoin on six benchmarks, including a semantic join benchmark [50], two benchmarks originally used for fuzzy-join evaluations [48, 75], and expanded variants of these benchmarks with larger search spaces, constructed by injecting tables from the WDC corpus [70]. To our knowledge, this is the first evaluation to directly assess the semantic alignment of values in columns retrieved by semantic join discovery methods. Thus, we adapt fuzzy-join benchmarks for this evaluation. First, we conduct extensive ablation studies to analyze different design choices of MosaicJoin. Next, we show that MosaicJoin achieves new state-of-the-art results on value-level semantic join search among published methods –outperforming the best-performing published method by up to 17.6% in NDCG@20 on smaller benchmarks and up to 135.7% in Precision@10 on larger benchmarks.

## 5.1 Setup

*5.1.1 Environment.* We implement MosaicJoin in Python using PyTorch and Huggingface transformers library. All experiments were conducted on a server equipped with a NVIDIA L40S GPU, 128GB RAM, and 8 CPU cores.

*5.1.2 Benchmarks.* We perform our experiments using three benchmarks and construct expanded variants of these with WDC tables (see details in Table 1). For each benchmark, we split tables into single-column tables to evaluate joinable column search.

**Auto-FuzzyJoin** [48] and **Webtables** [75] benchmarks were originally designed for evaluating joins between source and target tables. Auto-FuzzyJoin contains 50 diverse fuzzy-join table pairs derived from DBPedia snapshots [46] between 2013-2021, while Webtables consists of 31 table-pairs collected from Bing and Google search spanning 17 topics, where joins are transformation-based (e.g., formatting or lexical variations). Since both benchmarks evaluate value-level joins, we adapt them to our search setting. The queries are the annotated join columns from the source tables, and the search space consists of all columns across all tables. The ground truth is the corresponding annotated join column from the paired target table in the original ground truth.[2]

**Freyja** benchmark [50] was created for semantic and syntactic joinable column discovery. It comprises 160 datasets from Kaggle and OpenML, with manually labeled ground truth.[3]

To evaluate robustness at scale, we inject into all benchmarks a sample of the **WDC** Web Table Corpus [70], which has been previously used to evaluate table search [23, 31], into the data lakes of Auto-FuzzyJoin, Webtables, and Freyja.

*5.1.3 Baselines.* We compare our approach, MosaicJoin, with the state-of-the-art approaches for column-level join method [19], value-level method [18], proxy-based join method that bridges the two [31], semantic overlap set search methods [15, 54], and context-aware join method [41]. We use off-the-shelf model weights released by the authors for DeepJoin and Snoopy, and the fastText base model [45] for PEXESO. This evaluates all methods in the same setting, where they can be deployed directly on new or evolving data lakes without task-specific retraining, which is exactly the setting that MosaicJoin targets. To isolate value semantics and ensure a fair comparison across methods, we evaluate all methods using column values only, without column names. This values-only setting follows prior semantic join discovery evaluations [18, 31] and reflects a realistic data lake setting where metadata is often incomplete, noisy, or ambiguous [1, 25, 57].

• **DeepJoin** [19] is a column-level join discovery method that encodes each column into a single fixed-length embedding using SBERT. We use the model that the authors pre-trained on WDC [63] to run model inference on our benchmarks. To compare fairly with MosaicJoin, we only embed column values.[4]

• **PEXESO** [18], a value-level method, is the most similar baseline to MosaicJoin. PEXESO embeds all column values and retrieves candidates via pivot-based filtering and a grid index.[5]

[2] https://github.com/chu-data-lab/AutomaticFuzzyJoin/tree/master/src/autofj/benchmark,https://github.com/Yeye-He/Auto-Join/tree/master/autojoin-Benchmark
[3] https://freyja-data-discovery.github.io
[4] https://github.com/mutong184/deepjoin
[5] https://github.com/BIT-DataLab/LakeBench/tree/main/join/Pexeso

Table 1: Statistics on AUTO-FUZZYJOIN, WEBTABLES, and FREYJA benchmarks, and WDC tables used in the expanded benchmarks.

| Benchmark | Split | # Cols | Total # Rows | Max # Rows | Avg # Rows | Max Card. | Avg Card. |
|---|---|---|---|---|---|---|---|
| AUTO-FUZZYJOIN | Query | 50 | 164.7K | 6,933 | 3,295 | 6,933 | 3,295 |
| | Data Lake | 100 | 182K | 6,933 | 1,826 | 6,933 | 1,826 |
| WEBTABLES | Query | 32 | 3.8K | 690 | 120 | 673 | 113 |
| | Data Lake | 244 | 47K | 2,195 | 194 | 2,191 | 82 |
| FREYJA | Query | 50 | 223K | 57,793 | 4,459 | 1,303 | 159 |
| | Data Lake | 1,317 | 16M | 1,048,575 | 12,169 | 834,244 | 3,342 |
| WDC | Data Lake | 97,703 | 10M | 810 | 102 | 810 | 99 |

• **SNOOPY** [31] is a column-level method that captures value-level joinability using proxy columns, learned from the Scorpion model [31]. We use the original model weights trained on the WDC corpus.[6]
• **KOIOS** [54] is a value-level semantic join discovery method that embeds values and indexes them for neural similarity search. Since the original implementation is written in C, we re-implement it in Python while preserving the method's original maximum bipartite matching scoring logic to ensure a fair comparison.[7]
• **SILKMOTH** [15] is a value-level join discovery method based on set containment. It retrieves joinable columns by comparing the overlap between column value sets rather than semantic similarity. Since the original implementation is not publicly available, we implement a Python version of SILKMOTH according to the algorithms and specifications in the paper [15].
• **TOPJOIN** [41] is a hybrid join discovery method that combines value overlap, column-level semantic similarity, and table-context signals. We use the original implementation of TOPJOIN.[8]
• **EXACT-SEMANTIC-JOIN (ESJ)** is a variation of MOSAICJOIN that performs exact semantic join search without constructing semantic sketches of value embeddings proposed in Section 3.2. Instead, it computes exact semantic join score defined in Eq. (2) by, for each query value embedding, finding its closest value embedding in the candidate column (i.e., exact nearest-neighbor matching) and aggregating these matches at query time.

*5.1.4 Ground truth Expansion.* For the base benchmarks, we use the provided ground-truth labels. For the expanded benchmarks, no ground truth exists between the original query columns and the injected WDC tables, so we augment the original labels with silver labels generated by Gemini 2.5 Pro [11]. We evaluate using the union of the original benchmark ground truth and the silver labels. We use an LLM judge because many valid joins are semantic rather than exact equi-joins, and exhaustive value-pair annotation is infeasible at scale (e.g., FREYJA +WDC would require up to 20K × 26M value-pair comparisons). For each query–candidate pair, we provide both columns' unique values after minor string normalization and ask the judge to classify the pair as `equijoin`, `semantic`, or `not joinable`, with a confidence score and brief rationale. We retain only predictions with confidence > 0.7. To avoid bias toward MOSAICJOIN, for each query column we form the annotation pool as the union of the top-50 retrieved columns from MOSAICJOIN and each baseline, excluding pairs already present in the original ground truth. Thus, silver labels can originate from any method's retrieval list. To validate retained positive silver labels, we conduct a multi-model audit on a stratified sample of 200 joinable pairs across all expanded benchmarks, including 75 semantic joins and 125 equi-joins. Each pair is independently checked by Gemini 2.5 Pro, GPT-5.5 [60], and Gemini 3.5 Flash [30]; all 200 receive majority support, and 185 receive unanimous support. The remaining 15 disagreement cases are manually reviewed by three human curators and confirmed as correct joinable pairs. For context, with $n = 200$, a one-sided exact binomial test for detecting an improvement from 90% to 95% precision at $\alpha = 0.05$ has approximately 80% power, confirming the statistical significance of our verification results.

*5.1.5 Metrics.* Following previous work [19, 31], we report Recall@$k$ and NDCG@$k$ for AUTO-FUZZYJOIN and WT, since they expect a single joinable column per query column in their ground truth. For FREYJA, which expects multiple columns per query, we also evaluate Precision@$k$. For the expanded benchmarks, where we have LLM annotations of the top-10 retrieved columns from each method, we evaluate Precision@10 and NDCG@10.

For a query column $C_Q$, let $\mathcal{S} = \{C_1, \dots C_{|\mathcal{S}|}\}$ be the set of expected joinable columns and $\hat{\mathcal{S}} = \{\hat{C}_1, \dots \hat{C}_k\}$ be the set of retrieved top-$k$ joinable columns by a specific method with respect to $C_Q$. Precision@$k = \frac{|\mathcal{S} \cap \hat{\mathcal{S}}|}{|\hat{\mathcal{S}}|}$ and Recall@$k = \frac{|\mathcal{S} \cap \hat{\mathcal{S}}|}{|\mathcal{S}|}$. NDCG@$k$ measures the quality of the ranking of the top-$k$ retrieved columns. Since our ground-truth labels are binary, we define relevance as $\mathrm{rel}_i = \mathbf{1}\left[\hat{C}_i \in \mathcal{S}\right]$, where $\hat{C}_i$ is the column returned at rank $i$. NDCG@$k$ is defined as $\frac{DCG@k}{IDCG@k}$, where $DCG@k = \sum_{i=1}^{k} \frac{rel_i}{log_2(i+1)}$ and $IDCG@k = \sum_{i=1}^{min(k,|\mathcal{S}|)} \frac{1}{log_2(i+1)}$

For efficiency, we measure the query search time of all the methods. All metrics are averaged over all queries.

## 5.2 Ablation Study

We first conduct an ablation study of MOSAICJOIN 's design choices on the FREYJA benchmark to identify the best hyperparameter setting. For each ablation study, we keep the other hyperparameters fixed. We focus on this benchmark because it is technically our largest base benchmark and has the highest cardinality, measured by the number of rows with unique values (Table 1).

**Varying Size of Query Subsamples.** Figure 3 shows the NDCG@10 and runtime of MOSAICJOIN on the FREYJA benchmark as we vary the number $b$ of randomly sampled rows from each query column.

[6]https://github.com/ZJU-DAILY/Snoopy/tree/main
[7]https://github.com/DataIntelligenceCrew/koios-semantic-search
[8]https://github.com/IBM/ContextAwareJoin

**Table 2: Ablation studies of Precision, Recall, and NDCG for $k = 10, 20$ of MosaicJoin on Freyja for different sketch construction methods, online similarity methods, sketch sizes $m$, and embedding models. (*) denotes the hyperparameters we use in the rest of the experiments.**

| | Method | Precision@10 | Recall@10 | NDCG@10 | Precision@20 | Recall@20 | NDCG@20 |
|---|---|---|---|---|---|---|---|
| Sketch Construction | k-center* | 0.876 | 0.287 | 0.885 | 0.829 | 0.543 | 0.848 |
| | k-means | 0.454 | 0.143 | 0.482 | 0.259 | 0.163 | 0.335 |
| Online Similarity | chamfer ($\widehat{\mathcal{J}}_{\text{ch}}$) | 0.890 | 0.292 | 0.907 | 0.801 | 0.526 | 0.838 |
| | inverse chamfer ($\widehat{\mathcal{J}}_{\text{inv}}$) | 0.820 | 0.268 | 0.820 | 0.797 | 0.521 | 0.804 |
| | average chamfer ($\widehat{\mathcal{J}}_{\text{bi}}$)* | 0.876 | 0.287 | 0.885 | 0.829 | 0.543 | 0.848 |
| Sketch Sizes | $m = 128$ | 0.882 | 0.289 | 0.888 | 0.850 | 0.558 | 0.864 |
| | $m = 64$* | 0.876 | 0.287 | 0.885 | 0.829 | 0.543 | 0.848 |
| | $m = 32$ | 0.838 | 0.274 | 0.848 | 0.792 | 0.518 | 0.813 |
| Embedding Model | EmbeddingGemma [69] ($d = 128$)* | 0.876 | 0.287 | 0.885 | 0.829 | 0.543 | 0.848 |
| | MPNet [68] ($d = 768$) | 0.460 | 0.145 | 0.506 | 0.259 | 0.163 | 0.349 |
| | bge-base-en-v1.5 [71] ($d = 768$) | 0.906 | 0.297 | 0.909 | 0.892 | 0.585 | 0.899 |

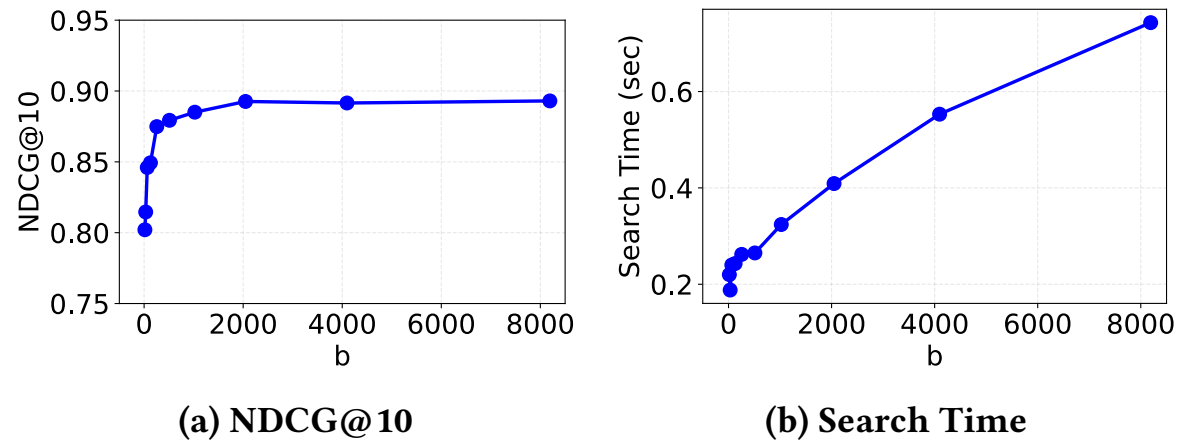

**(a) NDCG@10** **(b) Search Time**

**Figure 3: Varying $b$ (number of sampled values from query columns) on Freyja benchmark**

We observe that NDCG@10 quickly stabilizes around $b = 1024$, with little improvement beyond this point, while runtime is 4.5× faster than using the full query column. This indicates that random query subsampling preserves retrieval quality while improving efficiency. Based on this tradeoff, we use $b = 1024$ query samples in all experiments in Section 5.3. This empirical trend is consistent with the guarantee in Theorem 2 from Section 4.3.

**Comparison of Sketch Construction Methods** Table 2 shows the effect of different sampling strategies for constructing semantic sketches on the performance of MosaicJoin on Freyja. In particular, we compare our $k$-center-based method (Section 3.2) against a $k$-means clustering, where we use the standard $k$-means implementation from [62]. We observe that $k$-center sampling outperforms $k$-means by large margins for all metrics.

This difference is expected for two main reasons. First, $k$-means is optimized to minimize average squared distance to cluster centroids, which tends to bias the representation toward dense regions of the embedding space. On Freyja, where many values are repeated or differ only slightly, this can cause multiple centroids to concentrate around highly frequent but semantically similar values, leaving other parts of the column underrepresented. In contrast, $k$-center sampling explicitly aims to maximize coverage of the embedding space, producing a more diverse set of representative points. Second, $k$-means returns synthetic centroids that generally do not coincide with actual embedded values, whereas our $k$-center procedure selects representatives directly from the observed embedding set. As a result, the semantic sketch remains grounded in real points while still effectively covering the space via the farthest-first traversal used by $k$-center. Overall, these results show that $k$-center sampling yields a more faithful and diverse summary of the column embeddings, leading to better retrieval performance.

**Varying Online Similarity Methods.** To compare query columns against candidate sketches, we study several Chamfer-style similarity functions, following the same nearest-neighbor matching principle used in prior retrieval work such as Muvera [36]. All variants use dot-product similarity on normalized embeddings and differ only in how they aggregate these matches. This ablation helps determine which aggregation rule is best aligned with semantic join discovery.

*Directed Chamfer (query → candidate sketch).* Our primary score is the directed average Chamfer from the query to the candidate sketch:

$$\widehat{\mathcal{J}}_{\text{ch}}(C_Q, C) = \frac{1}{|\Phi(C_Q)|} \sum_{q \in \Phi(C_Q)} \max_{s \in S(C)} (q \cdot s).$$

This is the sketch-based approximation of the directed semantic joinability score from Section 3.1. It measures, for each query embedding, how well it is supported by the candidate sketch, and is therefore the most natural choice for our retrieval setting.

*Inverse Chamfer (candidate sketch → query).* We also evaluate the inverse direction:

$$\widehat{\mathcal{J}}_{\text{inv}}(C_Q, C) = \frac{1}{|S(C)|} \sum_{s \in S(C)} \max_{q \in \Phi(C_Q)} (s \cdot q).$$

Unlike the directed score above, this variant measures how well the candidate sketch is covered by the query. While this provides a meaningful alternative, it is less directly aligned with our objective, since in join retrieval we primarily care about whether query values can be matched by the candidate column.

*Bidirectional (average) Chamfer.* Finally, we consider a symmetric variant obtained by averaging the two directions: $\widehat{\mathcal{J}}_{\text{bi}}(C_Q, C) = \frac{1}{2}\left(\widehat{\mathcal{J}}_{\text{ch}}(C_Q, C) + \widehat{\mathcal{J}}_{\text{inv}}(C_Q, C)\right)$. This score balances both notions of coverage, rewarding pairs that match well in both directions. It can be viewed as a more symmetric measure of semantic similarity,

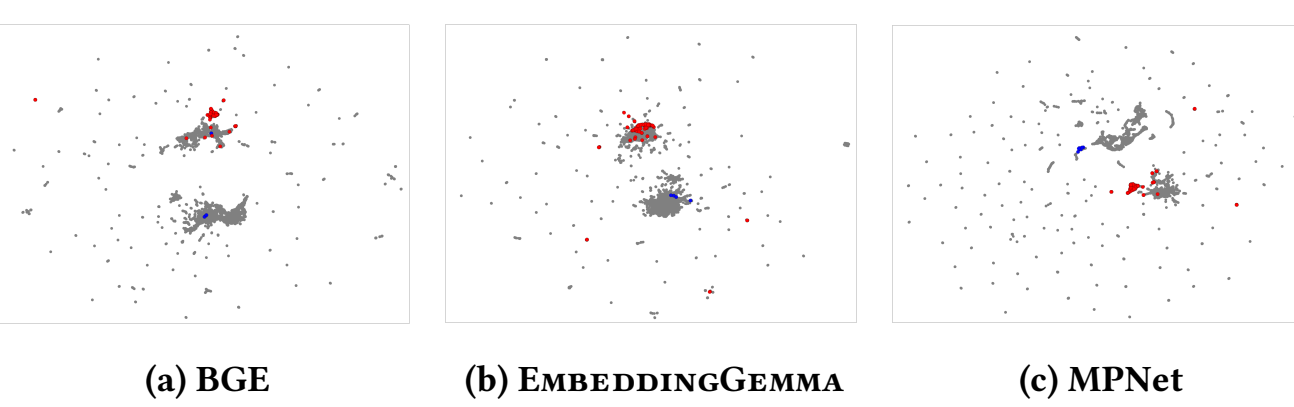

(a) BGE (b) EmbeddingGemma (c) MPNet

**Figure 4: UMAP projections of Freyja values embeddings from BGE, EmbeddingGemma, and MPNet Models. All embeddings are in gray. Red and blue points represent expected column values for two distinct queries.**

though such symmetry is not necessarily required in our inherently directed retrieval task.

In all three cases, we use *averages* to keep scores comparable across query columns with different numbers of values and candidate sketches of different sizes. As shown in Table 2, directed Chamfer and average Chamfer perform comparably. Since the difference is marginal, we adopt average Chamfer in all subsequent experiments and omit the comparison between direct Chamfer against the baselines due to space constraints.

**Varying Sketch Size $m$:** We next study the effect of the sketch size parameter $m$ in the $k$-center sketch construction algorithm (Algorithm 1) on the Freyja benchmark. Here, $m$ controls the number of representative embeddings retained in each semantic sketch. As shown in Table 2, increasing $m$ beyond 64 yields only marginal gains in retrieval performance. At the same time, smaller sketches reduce search cost (e.g., $m = 64$ is 1.25× faster than $m = 128$), leading to faster query-time performance. Based on this tradeoff, we use $m = 64$ in our experiments. More generally, this accuracy-efficiency balance can be tuned depending on system requirements, as further illustrated in Figure 1.

**Varying Embedding Models:** We also compare different embedding models for constructing value representations: EmbeddingGemma [69], BGE [71], and MPNet [68]. We analyze the underlying structure of the embedding space. Figure 4 shows a uniform manifold approximation and projection (UMAP) visualization [51] of all embeddings of Freyja values (shown in gray) and highlights the distribution of the expected columns for two sample queries in red and blue. To evaluate embedding quality, we compute the average cosine distance of values in each expected column to its respective group centroid. Smaller cosine distance signals compact clusters, meaning the embedding model has successfully captured the shared semantic context of a column and differentiated it from others. BGE, EmbeddingGemma, and MPNet embeddings have average cosine distances of 0.087, 0.133, and 0.174, respectively. The relative ranking of these models by cluster compactness (BGE, EmbeddingGemma, MPNet, in descending order) aligns exactly with their relative ranking in join discovery performance (Table 2). Thus, the embedding quality directly impacts the downstream performance.

While BGE [71] with output dimension 768 achieves slightly higher retrieval accuracy, it incurs a substantially higher query-time cost: its average search time is 0.931 seconds per query, compared to only 0.080 seconds for EmbeddingGemma [69] with output dimension 128. MPNet [68] is also fast in search time (0.098 seconds),

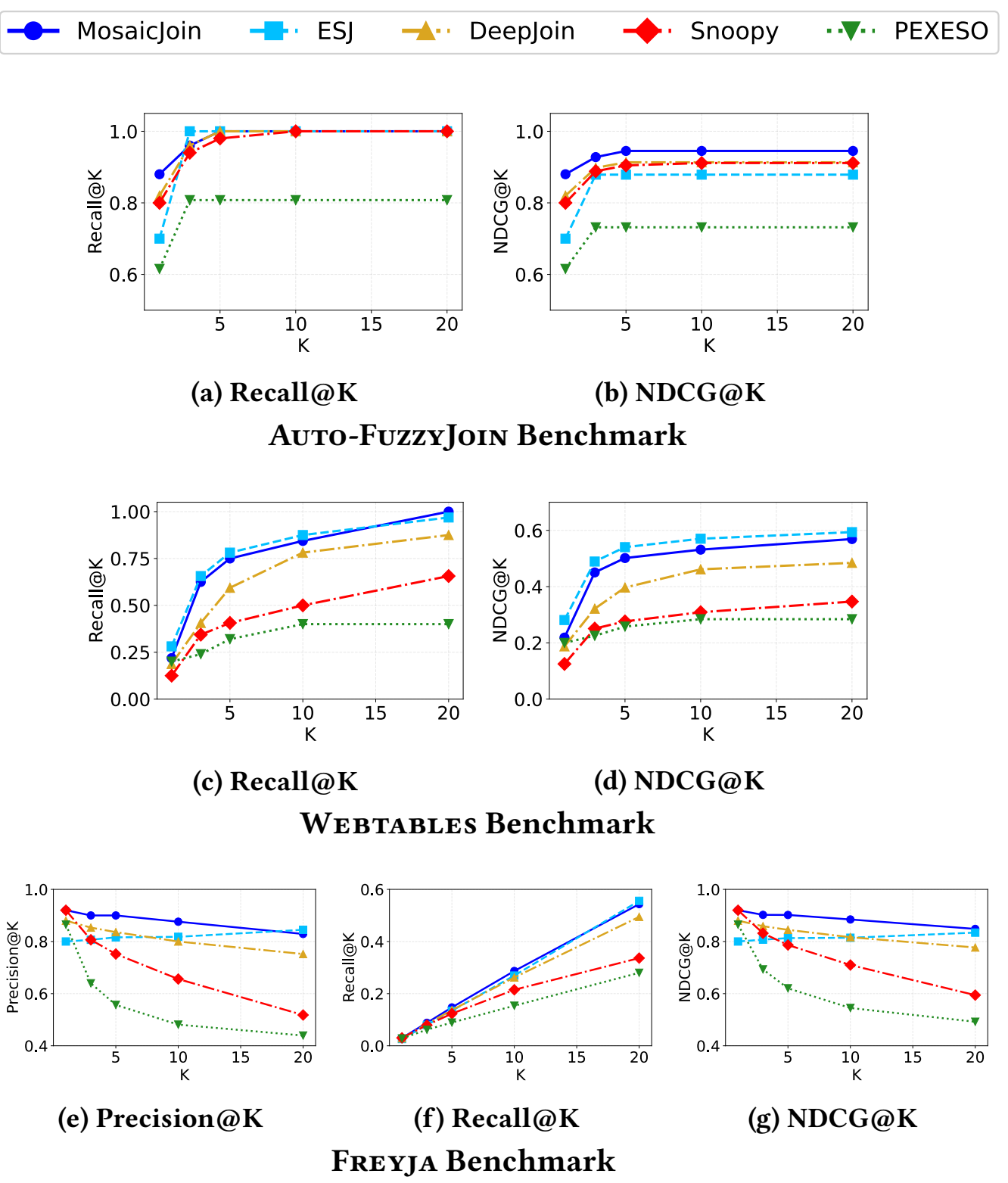

(a) Recall@K (b) NDCG@K
Auto-FuzzyJoin Benchmark

(c) Recall@K (d) NDCG@K
Webtables Benchmark

(e) Precision@K (f) Recall@K (g) NDCG@K
Freyja Benchmark

**Figure 5: Effectiveness of MosaicJoin and baselines for varying $k$'s on Auto-FuzzyJoin, Webtables, and Freyja.**

but has significantly lower accuracy performance. We therefore select EmbeddingGemma as the default embedding model because it offers a much stronger efficiency–accuracy tradeoff.

A key reason this works well is that EmbeddingGemma is trained with Matryoshka Representation Learning [44], which concentrates the most informative semantic signal in the leading dimensions. As a result, truncating the representation to 128 dimensions preserves similar retrieval quality as higher-dimensional embeddings while significantly reducing storage and similarity computation costs at query time. Based on this tradeoff, we use EmbeddingGemma with 128-dimensional outputs in subsequent experiments.

## 5.3 Comparisons against Baselines

Table 3 reports the effectiveness results on all base and expanded benchmarks. MosaicJoin outperforms existing methods DeepJoin, Snoopy, PEXESO on all six benchmarks. We also compare against ESJ, our own exact-search variant, in detail later in this section.

**Base Benchmarks.** On Freyja, MosaicJoin outperforms DeepJoin, Snoopy, PEXESO by 9.1%, 42.8%, 72.4% in NDCG@20, respectively, and by 10.2%, 60%, 88.8% in Precision@20. On Auto-FuzzyJoin and Webtables, MosaicJoin outperforms the best-performing baseline, DeepJoin, by 3.5% and 17.6% in NDCG@20, respectively. As shown in Figure 5, MosaicJoin consistently outperforms the baselines for all values of $k$ on all benchmarks.

Since DeepJoin collapses each query column into a single fixed-length embedding, it cannot capture fine-grained value alignment,

Table 3: Effectiveness of MosaicJoin and external baselines DeepJoin, Snoopy, and PEXESO, and our exact-search variant ESJ on all benchmarks. PEXESO timed out for most, if not all, queries on the expanded benchmarks.

| Method | Auto-FuzzyJoin | | WebTables | | Freyja | | | Auto-FuzzyJoin +WDC | | WebTables +WDC | | Freyja +WDC | |
|---|---|---|---|---|---|---|---|---|---|---|---|---|---|
| | R@20 | NDCG@20 | R@20 | NDCG@20 | P@20 | R@20 | NDCG@20 | P@10 | NDCG@10 | P@10 | NDCG@10 | P@10 | NDCG@10 |
| ESJ | 1.000 | 0.879 | 0.969 | 0.594 | 0.845 | 0.556 | 0.834 | 0.526 | 0.716 | 0.072 | 0.462 | 0.428 | 0.625 |
| DeepJoin | 1.000 | 0.913 | 0.875 | 0.484 | 0.752 | 0.495 | 0.777 | 0.422 | 0.678 | 0.025 | 0.141 | 0.390 | 0.484 |
| Snoopy | 1.000 | 0.911 | 0.656 | 0.347 | 0.518 | 0.336 | 0.594 | 0.494 | 0.778 | 0.028 | 0.146 | 0.330 | 0.506 |
| PEXESO | 0.808 | 0.732 | 0.400 | 0.284 | 0.439 | 0.280 | 0.492 | – | – | – | – | – | – |
| MosaicJoin | 1.000 | 0.945 | 1.000 | 0.569 | 0.829 | 0.543 | 0.848 | 0.630 | 0.819 | 0.066 | 0.423 | 0.428 | 0.562 |

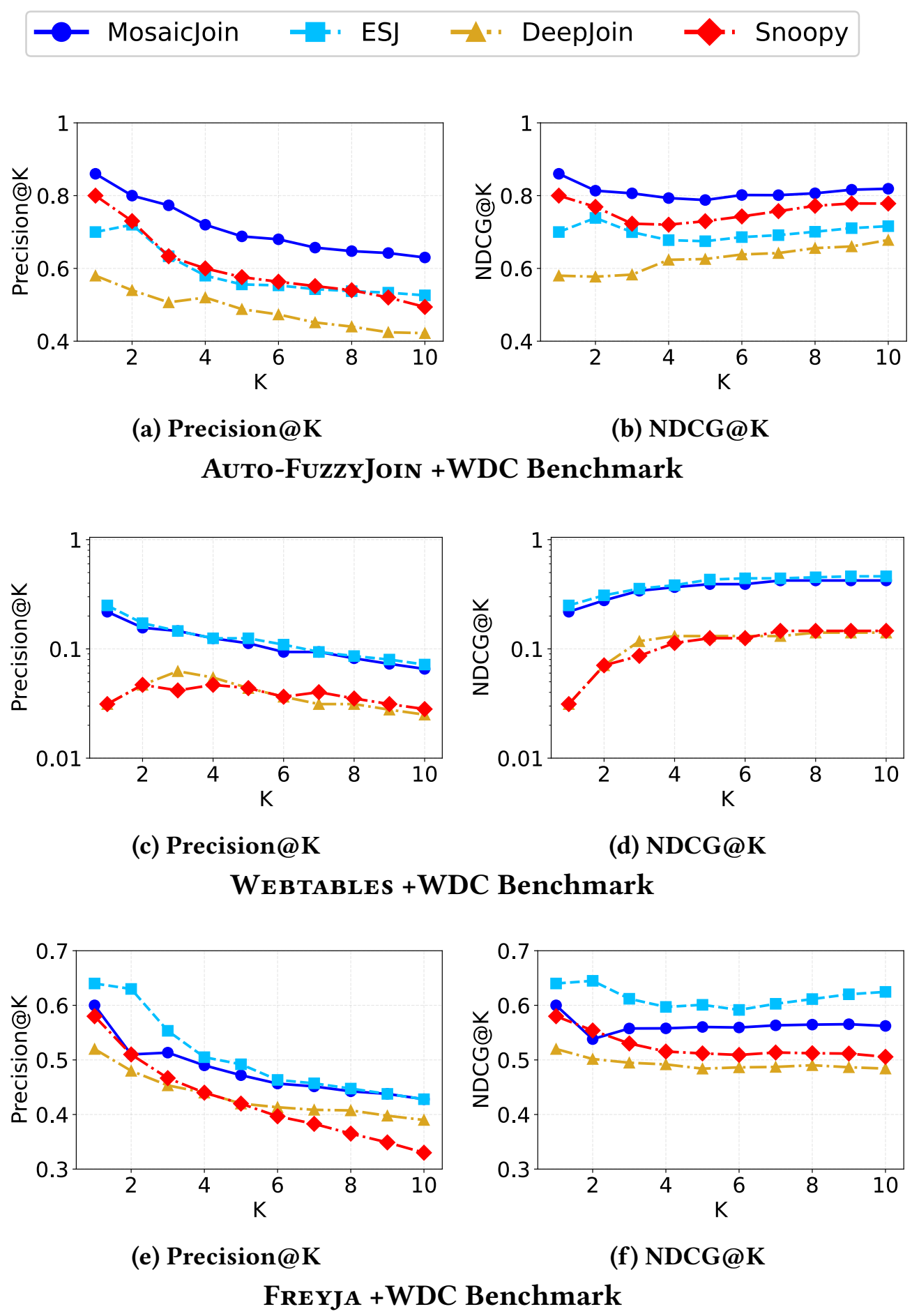

Figure 6: Effectiveness of MosaicJoin and baselines for varying $k$'s on the expanded benchmarks.

thus producing false positives when a candidate column has a similar topic but is not value-level joinable. In contrast, Snoopy projects columns onto learned proxy columns to determine column joinability and to capture fine-grained value alignment. However, its proxies are trained on the WDC corpus, which may not capture the value distributions in our benchmark, including the long-tail distributions in high-cardinality columns such as those in Freyja. On the base benchmarks, DeepJoin actually outperforms Snoopy because DeepJoin's SBERT encoder generalizes better out-of-domain. However, on the expanded benchmarks that include injected tables from WDC (as shown in Figure 6), Snoopy's specialized in-domain proxies cover the data distributions, leading to better performance than DeepJoin's. Finally, although PEXESO is value-level, its accuracy is low because its pivot-based pruning can prune many joinable columns, and by counting the number of query vectors having at least one matching vector above a fixed similarity threshold, it can return many false positives. In contrast, MosaicJoin's sketches and Chamfer scoring preserve value-level semantic coverage without relying on similarity thresholds or training data.

**Expanded Benchmarks.** After injecting WDC tables into Auto-FuzzyJoin, WebTables, Freyja benchmarks, we evaluate Precision@10 and NDCG@10 on the expanded benchmarks using LLM annotations. PEXESO times out after 10 minutes per query, and thus, we cannot report its results for the expanded benchmarks. On Freyja +WDC, MosaicJoin outperforms DeepJoin and Snoopy by 9.7%, 29.7% in Precision@10, respectively. On Auto-FuzzyJoin +WDC and WebTables +WDC, MosaicJoin outperforms the best performing baseline, Snoopy, by 27.5% and 135.7% in Precision@10, respectively. The absolute scores are lower on WebTables +WDC for all methods because most of injected WDC columns are not joinable with WebTables queries (72% of columns), thus resulting in many false positives in top-$k$. Figure 6 shows that MosaicJoin outperforms existing methods throughout all values of $k$. Thus, MosaicJoin is robust to larger search spaces, still outperforming existing methods on value-driven semantic join discovery.

**Comparison with ESJ.** As shown in Table 3, the exact variant of MosaicJoin, ESJ, outperforms MosaicJoin by 4.4% on WebTables, by 9.2% on WebTables +WDC in NDCG@$k$, and by 11.2% on Freyja +WDC. Since ESJ performs exhaustive pairwise matching over all value embeddings, it captures every possible value alignment, including value alignments that MosaicJoin's sketches may miss. Meanwhile, we observe the opposite trend on Auto-FuzzyJoin and Auto-FuzzyJoin +WDC, where MosaicJoin outperforms ESJ by 7.5% and 14.4% in NDCG@$k$, respectively. Auto-FuzzyJoin contains one-to-many value alignments [48], and ESJ's exhaustive matching is sensitive to redundant or noisy variations of the same entity. In contrast, MosaicJoin's $k$-center sketching selects distinct representative values, thus handling these complex alignments more robustly than the exhaustive variant.

**Robustness to Query Column Size.** We further analyze how query column size affects Precision@10, Recall@10, and NDCG@10 of all methods on the Freyja benchmark, which contains the largest query columns in our evaluation (up to 57K values). We evenly split the query columns into five groups of varying numbers of rows.

As shown in Figure 7, MosaicJoin and ESJ are robust as the query size grows to 57K, while the performances of DeepJoin,

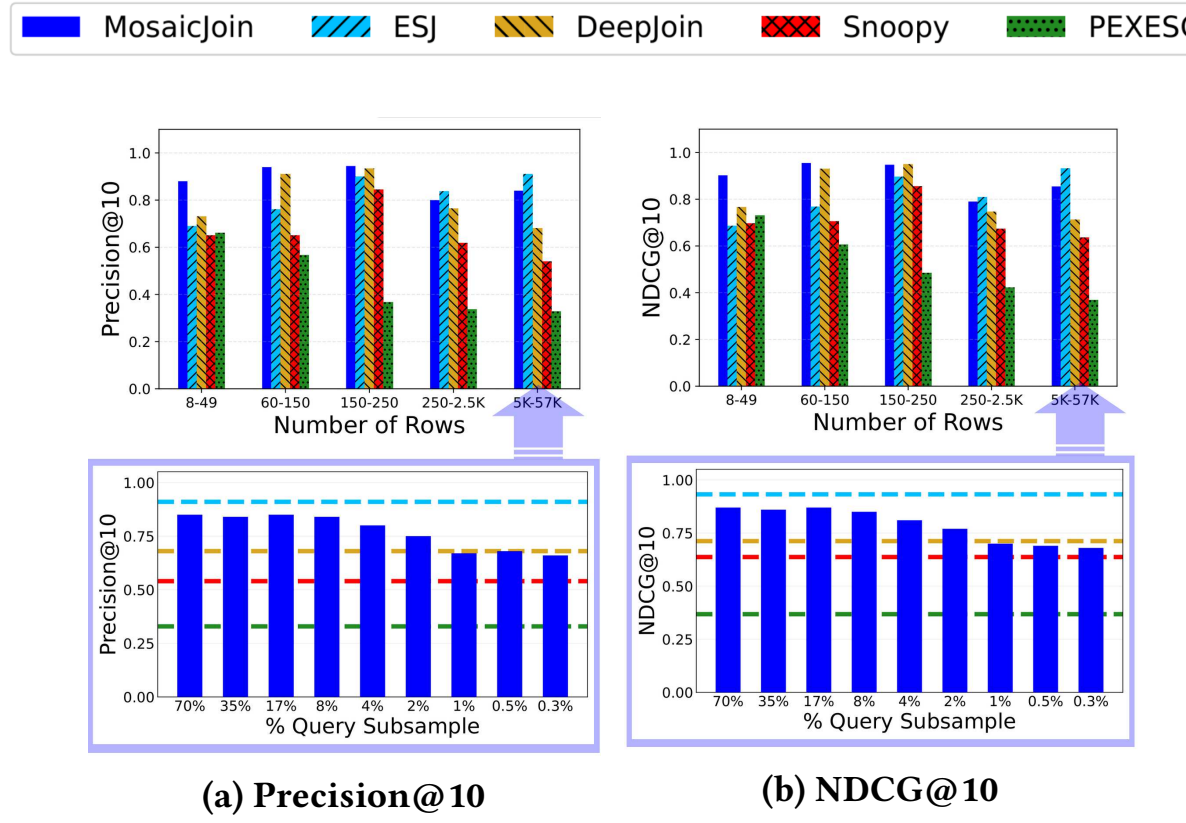

**(a) Precision@10** **(b) NDCG@10**

**Figure 7: Analysis of Precision@10 and NDCG@10 of MosaicJoin, ESJ, DeepJoin, Snoopy, PEXESO as we vary the number of rows in query columns in the Freyja benchmark. Recall@10 follows a similar trend. The bottom row subsamples values from the largest query bucket (up to 57K rows) in MosaicJoin; dashed lines show the performance of the baselines.**

Snoopy, and PEXESO gradually decrease. Among the baselines, PEXESO shows the lowest overall performance. Since PEXESO embeds all query values and relies on pivot-based filtering to prune candidates, it becomes increasingly expensive and less effective as the query column size increases. DeepJoin's performance degrades more gradually as the query size increases. DeepJoin truncates columns to fit its models' context windows, so the performance is stable on smaller queries but declines as DeepJoin loses potentially joinable values as the query size increases to 57K. Snoopy's performance is generally lower than DeepJoin's, since its proxies are trained on WDC. As the query column size increases, Snoopy's performance declines since its fixed-size proxies fail to represent the long-tail of large value distributions.

On the other hand, MosaicJoin and ESJ maintain stable performance across queries of all sizes. ESJ computes Chamfer score exhaustively over all query value embeddings, so its accuracy remains high. However, its runtime grows linearly with query size, taking ~16 seconds per query on Freyja. In contrast, the $k$-center sketches in MosaicJoin preserve the semantic coverage of candidate columns, regardless of their size (detailed in Section 3.2), while the query subsampling bounds the online cost of scoring regardless of query size (Theorem 2), leading to accuracy that is similar to ESJ at a faster search time of ~0.4 seconds. We see that MosaicJoin achieves high accuracy at lower query cost, even on large query columns, where all published baselines' performances degrade.

To evaluate robustness as query columns scale, we vary the percentage of $b$ sampled values on the largest query columns in the Freyja benchmark (5K to 57K rows). By reducing $b$ on fixed column sizes, this experiment serves as a controlled proxy for scaling beyond Freyja benchmark size as it represents the same reduction in sampling ratio as increasing the column size for a fixed $b$. As shown on the bottom of Figure 7, as the query sampling ratio decreases for larger columns, there is a slight decline in accuracy. However, even when sampling less than 2% of query values, MosaicJoin performs comparably to DeepJoin, Snoopy, and PEXESO. While sampling does impact accuracy, these results demonstrate that MosaicJoin remains relatively robust as query columns grow.

**Efficiency.** Table 4 reports the preprocessing overhead for MosaicJoin and ESJ on Freyja, which is the largest benchmark. While MosaicJoin generates both embeddings and sketches and ESJ only uses embeddings, the additional time and storage overhead for sketches is minimal. Specifically, sketch creation adds only 0.08, 0.07, and 0.25 seconds per column on Auto-FuzzyJoin, Webtables, and Freyja benchmarks, respectively. Sketch construction does not increase peak memory requirements: Peak RAM and Peak GPU Memory for both MosaicJoin and ESJ are 3.69 GB and 25.71 GB, respectively. Importantly, this preprocessing is a one-time offline cost that scales linearly through parallelization across columns. When a new column is ingested, the update cost involves generating its embedding and sketch. For the Freyja benchmark, this update cost is 2.48 seconds and requires ~52 MB of storage. Since columns are processed independently, ingesting new columns does not require reprocessing.

Table 5 reports the search time of all methods, averaged over all queries, on Auto-FuzzyJoin, Webtables, and Freyja. MosaicJoin is 2 to 66 times faster than the existing value-level joinable column baseline, PEXESO. Compared to the exact semantic search baseline ESJ, MosaicJoin is 2 to 49 times faster while achieving comparable or higher accuracy. Compared to overlap set search methods, MosaicJoin is 6 to 19 times faster than KOIOS and 120 to 224 times faster than SILKMOTH. SILKMOTH times out on the Freyja benchmark after 10 minutes per query. Although KOIOS uses a FAISS embedding index and a filter-verification framework, the bipartite matching used in KOIOS scoring makes the search slower than the Chamfer-style scoring in MosaicJoin. However, KOIOS is faster than ESJ on the Freyja benchmark since ESJ must exhaustively compare all embeddings for large queries while KOIOS effectively prunes the search space. MosaicJoin is also up to 1.4 times faster than TOPJoin, which uses column metadata and aggregate statistics and value profiles to retrieve joinable columns. As expected, single-vector methods like DeepJoin and Snoopy are faster than MosaicJoin at query time, since they compare one vector per candidate column rather than $m = 64$ sketch vectors. However, Figure 1 shows that the fast search time of these methods also results in lower accuracy. Beyond search time, DeepJoin and Snoopy require expensive offline training on large corpora, whereas MosaicJoin requires no training, making it easier to deploy on new data.

During online search on the largest benchmark Freyja, MosaicJoin and ESJ require 27% and 49% peak GPU Utilization, respectively, to embed query columns and compute Chamfer-style scores. MosaicJoin requires 2.39 GB of peak RAM and 2.13 GB of peak GPU memory, while ESJ requires 9.06 GB of peak RAM and 6.09 GB of peak GPU memory. These metrics demonstrate that the compact sketches in MosaicJoin not only save memory but also significantly reduce the computational load on the GPU compared to ESJ. In contrast, the value-level baseline PEXESO requires 3% of GPU utilization, 8.53 GB of peak RAM and 0.51 GB of peak GPU memory, since it mainly uses CPU for pruning and verification. However, MosaicJoin achieves faster search time and nearly four times lower memory usage than ESJ and PEXESO.

**Indexing.** We further evaluate MosaicJoin against baseline methods that use indexing, specifically overlap set search methods KOIOS

**Table 4: Preprocessing and update costs of MosaicJoin and ESJ on the Freyja benchmark.**

| | Time / Col | Total Time | Storage / Col | Total Storage |
|---|---|---|---|---|
| MosaicJoin | 2.48 sec | 3,259.17 sec | 52.17 MB | 68.65 GB |
| ESJ | 2.23 sec | 2,935.38 sec | 51.96 MB | 68.38 GB |

**Table 5: Average Online Search Time (in seconds) on Auto-FuzzyJoin, Webtables, Freyja benchmarks.**

| Method | Auto-FuzzyJoin | Webtables | Freyja |
|---|---|---|---|
| DeepJoin | 0.002 | 0.001 | 0.001 |
| Snoopy | 0.004 | 0.001 | 0.001 |
| PEXESO | 5.256 | 0.071 | 1.337 |
| ESJ | 0.507 | 0.060 | 15.646 |
| KOIOS | 1.547 | 0.383 | 1.832 |
| SILKMOTH | 17.936 | 4.809 | – |
| TOPJoin | 0.106 | 0.040 | 0.459 |
| MosaicJoin | 0.080 | 0.040 | 0.322 |

and SILKMOTH, and TOPJoin. To allow for a fair comparison of scoring methods, we incorporate the same Faiss embedding index used in KOIOS for MosaicJoin candidate retrieval. Figure 8 shows the effectiveness results on the Auto-FuzzyJoin benchmark, which is the largest benchmark where all methods successfully complete within a 10-minute query timeout.[9] Additionally, Auto-FuzzyJoin contains one-to-many column cardinalities, which is a realistic scenario in a data lake setting where denormalized tables often result in multiple values mapping to a single query value [48]. As shown in Figure 8, MosaicJoin outperforms TOPJoin, KOIOS, and SILKMOTH by 8.7%, 15.1%, and 523.4% in NDCG@20, respectively. Thus, Chamfer-style joinability in MosaicJoin is better suited for the one-to-many semantic mappings prevalent in noisy data lakes compared to exact bipartite matching.

We further analyze the distinction between value-level methods like MosaicJoin and hybrid methods like TOPJoin. MosaicJoin successfully finds semantically joinable columns that contain semantically aligned values. For example, in the Auto-FuzzyJoin benchmark, query values like "2004 Tiger Cup" and joinable column values like "2004 AFF Championship" share no string overlap, but both refer to the same sports tournament. In this case, MosaicJoin ranks this column first, whereas TOPJoin ranks it below top 3. TOPJoin utilizes table-level context, including column headers and table metadata, which is helpful when value semantics and thus the joins are ambiguous. For example, for a query column "country" in table "USA cars datasets" with values {"usa", "canada"}, TOPJoin successfully retrieves a joinable column on "sales territory" with values {"Northwest", "Northeast"}. While MosaicJoin currently relies on value-level information, we will explore incorporating table-level contextual signals in future work.

## 6 CONCLUSION

We present MosaicJoin, a value-driven semantic join discovery method that achieves the high accuracy of value-level methods at much faster query search time. MosaicJoin constructs $k$-center semantic sketches for each data lake column that maximize coverage of the column's embedding space, producing a compact representation that captures the full semantics of even high-cardinality columns and enables Chamfer joinability scoring at a cost bounded by the sketch size, rather than by the column size. A query subsampling estimator further reduces online query search time regardless of query column size, while preserving accuracy within provable error bounds, enabling robust retrieval even for query columns with up to 57K values. Experiments on fuzzy-join and semantic-join benchmarks, including larger variants, show that MosaicJoin outperforms existing methods across all benchmarks, making it immediately deployable on new data without task-specific retraining. However, value-level join discovery methods involve an inherent trade-off between accuracy and efficiency, as shown in Figure 1. Although MosaicJoin achieves sub-second retrieval, there is a trade-off between sketch size and retrieval accuracy, in which a larger sketch size may result in higher accuracy but slower runtime. If we consider all value embeddings in a column, like in ESJ, the accuracy may be higher (Figures 5 and 6) but the runtime is much slower (Table 5). Our results across benchmarks and ablation studies (Table 2) show that the accuracy of this sketch-based method can vary depending on the embedding quality, sketch size, and the sketch construction method. Furthermore, while query subsampling significantly reduces search time (Figure 3), especially for larger columns in the Freyja benchmark, subsampling may overlook rare joinable values in extremely skewed distributions. While we focus solely on values to find semantic joins, in future work we will explore whether incorporating metadata, semantic type annotations, and column profiles [28, 42, 72] could improve precision by capturing column-level and table-level context to better disambiguate semantically similar columns [40]. Additionally, while MosaicJoin's query search time is already faster than other value-level methods, even for benchmarks with up to 99K columns that each contain up to 1M values, approximate nearest neighbor indexing over sketch embeddings [36] could enable faster retrieval for very large data lakes.

[9] Experiments on Webtables and Freyja are included in the technical report [24].

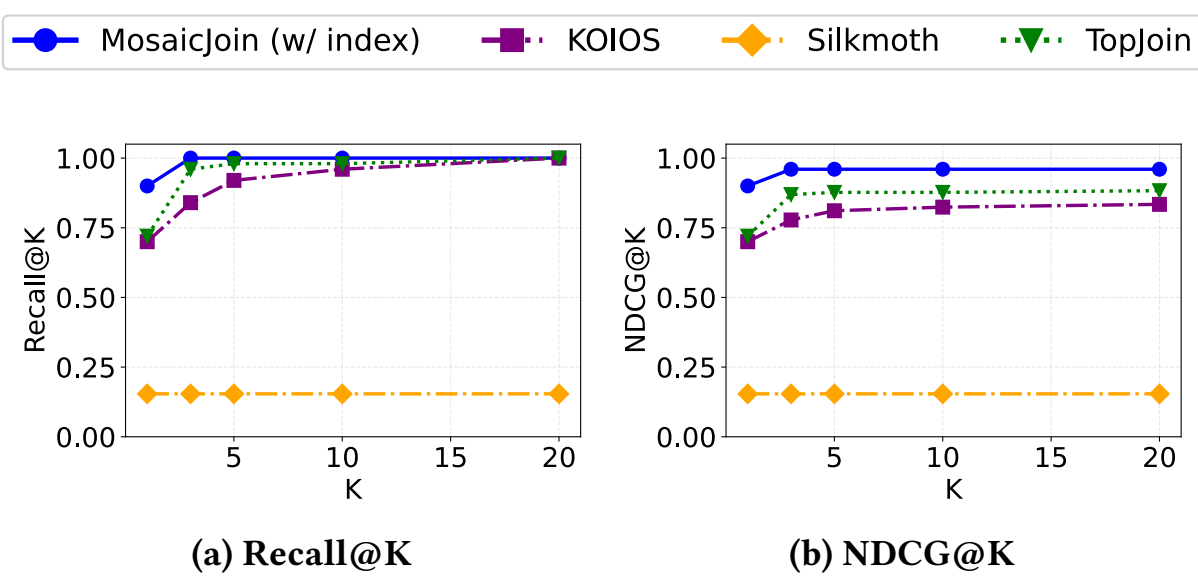

(a) Recall@K (b) NDCG@K

**Figure 8: Effectiveness of MosaicJoin with an index on Auto-FuzzyJoin, compared to KOIOS, SILKMOTH, TOPJoin.**

## ACKNOWLEDGMENTS

This work was supported in part by DARPA ASKEM (HR0011262087), ARPA-H BDF, and NSF (OAC-2411221). The views, opinions, and findings expressed are those of the authors and should not be interpreted as representing the views or policies of these agencies.

# REFERENCES

[1] Marco D. Adelfio and Hanan Samet. 2013. Schema Extraction for Tabular Data on the Web. *Proc. VLDB Endow.* 6, 6 (2013), 421–432.

[2] Aline Bessa, Juliana Freire, Tamraparni Dasu, and Divesh Srivastava. 2020. Effective discovery of meaningful outlier relationships. *ACM Transactions on Data Science* 1, 2 (2020), 1–33.

[3] Alex Bogatu, Alvaro A. A. Fernandes, Norman W. Paton, and Nikolaos Konstantinou. 2020. Dataset Discovery in Data Lakes. In *ICDE*. 709–720.

[4] Dan Brickley, Matthew Burgess, and Natasha F. Noy. 2019. Google Dataset Search: Building a search engine for datasets in an open Web ecosystem. In *WWW*. 1365–1375.

[5] Michael J. Cafarella, Alon Y. Halevy, and Nodira Khoussainova. 2009. Data Integration for the Relational Web. *Proc. VLDB Endow.* 2, 1 (2009), 1090–1101.

[6] Riccardo Cappuzzo, Aimee Coelho, Félix Lefebvre, Paolo Papotti, and Gaël Varoquaux. 2025. Retrieve, Merge, Predict: Augmenting Tables with Data Lakes. *TMLR* (2025).

[7] Sonia Castelo, Rémi Rampin, Aécio S. R. Santos, Aline Bessa, Fernando Chirigati, and Juliana Freire. 2021. Auctus: A Dataset Search Engine for Data Discovery and Augmentation. *Proc. VLDB Endow.* 14, 12 (2021), 2791–2794.

[8] Adriane Chapman, Elena Simperl, Laura Koesten, George Konstantinidis, Luis-Daniel Ibáñez, Emilia Kacprzak, and Paul Groth. 2020. Dataset search: a survey. *VLDB J.* 29, 1 (2020), 251–272.

[9] Nadiia Chepurko, Ryan Marcus, Emanuel Zgraggen, Raul Castro Fernandez, Tim Kraska, and David R. Karger. 2020. ARDA: Automatic Relational Data Augmentation for Machine Learning. *Proc. VLDB Endow.* 13, 9 (2020), 1373–1387.

[10] Fernando Chirigati, Harish Doraiswamy, Theodoros Damoulas, and Juliana Freire. 2016. Data polygamy: the many-many relationships among urban spatio-temporal data sets. In *SIGMOD*. 1011–1025.

[11] Gheorghe Comanici, Eric Bieber, Mike Schaekermann, Ice Pasupat, Noveen Sachdeva, Inderjit Dhillon, Marcel Blistein, Ori Ram, Dan Zhang, Evan Rosen, et al. 2025. Gemini 2.5: Pushing the frontier with advanced reasoning, multimodality, long context, and next generation agentic capabilities. *arXiv preprint arXiv:2507.06261* (2025).

[12] Tianji Cong, James Gale, Jason Frantz, H. V. Jagadish, and Çagatay Demiralp. 2023. WarpGate: A Semantic Join Discovery System for Cloud Data Warehouses. In *CIDR*. 1–7.

[13] Tianji Cong, Fatemeh Nargesian, and HV Jagadish. 2023. Pylon: Semantic Table Union Search in Data Lakes. *arXiv preprint arXiv:2301.04901* (2023).

[14] Arash Dargahi Nobari and Davood Rafiei. 2024. DTT: An example-driven tabular transformer for joinability by leveraging large language models. *Proc. ACM Manag. Data* 2, 1 (2024), 1–24.

[15] Dong Deng, Albert Kim, Samuel Madden, and Michael Stonebraker. 2017. SilkMoth: An Efficient Method for Finding Related Sets with Maximum Matching Constraints. *Proc. VLDB Endow.* 10, 10 (2017), 1082–1093.

[16] Yuhao Deng, Chengliang Chai, Lei Cao, Qin Yuan, Siyuan Chen, Yanrui Yu, Zhaoze Sun, Junyi Wang, Jiajun Li, Ziqi Cao, et al. 2024. Lakebench: A benchmark for discovering joinable and unionable tables in data lakes. *Proc. VLDB Endow.* 17, 8 (2024), 1925–1938.

[17] Yuyang Dong and Masafumi Oyamada. 2022. Table enrichment system for machine learning. In *SIGIR*. 3267–3271.

[18] Yuyang Dong, Kunihiro Takeoka, Chuan Xiao, and Masafumi Oyamada. 2021. Efficient joinable table discovery in data lakes: A high-dimensional similarity-based approach. In *ICDE*. IEEE, 456–467.

[19] Yuyang Dong, Chuan Xiao, Takuma Nozawa, Masafumi Enomoto, and Masafumi Oyamada. 2023. DeepJoin: Joinable Table Discovery with Pre-Trained Language Models. *Proc. VLDB Endow.* 16, 10 (2023), 2458–2470.

[20] Mahdi Esmailoghli, Jorge-Arnulfo Quiané-Ruiz, and Ziawasch Abedjan. 2022. MATE: Multi-Attribute Table Extraction. *Proc. VLDB Endow.* 15, 8 (2022), 1684–1696.

[21] Grace Fan and Juliana Freire. 2025. Hierarchical table semantics for exploratory table discovery. In *Proceedings of the Workshop on Human-In-the-Loop Data Analytics*. 1–7.

[22] Grace Fan, Jin Wang, Yuliang Li, and Renée J. Miller. 2023. Table Discovery in Data Lakes: State-of-the-art and Future Directions. In *SIGMOD Conference Companion*. ACM, 69–75.

[23] Grace Fan, Jin Wang, Yuliang Li, Dan Zhang, and Renée J. Miller. 2023. Semantics-aware Dataset Discovery from Data Lakes with Contextualized Column-based Representation Learning. *Proc. VLDB Endow.* 16, 7 (2023), 1726–1739.

[24] Grace Fan, Eden Wu, Majid Daliri, and Juliana Freire. 2026. Technical Report on MosaicJoin: Compact Semantic Sketches for Value-Level Join Discovery. https://github.com/gracefan2020/MosaicJoin/blob/main/technical_report.pdf

[25] Mina H. Farid, Alexandra Roatis, Ihab F. Ilyas, Hella-Franziska Hoffmann, and Xu Chu. 2016. CLAMS: Bringing Quality to Data Lakes. In *SIGMOD*. 2089–2092.

[26] Raul Castro Fernandez, Ziawasch Abedjan, Famien Koko, Gina Yuan, Samuel Madden, and Michael Stonebraker. 2018. Aurum: A Data Discovery System. In *ICDE*. 1001–1012.

[27] Raul Castro Fernandez, Jisoo Min, Demitri Nava, and Samuel Madden. 2019. Lazo: A cardinality-based method for coupled estimation of jaccard similarity and containment. In *ICDE*. 1190–1201.

[28] Benjamin Feuer, Yurong Liu, Chinmay Hegde, and Juliana Freire. 2024. ArcheType: A Novel Framework for Open-Source Column Type Annotation using Large Language Models. *Proc. VLDB Endow.* 17, 9 (2024), 2279–2292.

[29] Juliana Freire, Grace Fan, Benjamin Feuer, Christos Koutras, Yurong Liu, Eduardo Pena, Aécio Santos, Cláudio T Silva, and Eden Wu. 2025. Large language models for data discovery and integration: Challenges and opportunities. *IEEE Data Engineering Bulletin* (2025).

[30] Google. 2026. Gemini 3.5 Flash. https://ai.google.dev/gemini-api/docs/models/gemini-3.5-flash.

[31] Yuxiang Guo, Yuren Mao, Zhonghao Hu, Lu Chen, and Yunjun Gao. 2025. Snoopy: Effective and Efficient Semantic Join Discovery via Proxy Columns. *IEEE Trans. Knowl. Data Eng.* 37, 5 (2025), 2971–2985.

[32] Alon Y. Halevy, Flip Korn, Natalya Fridman Noy, Christopher Olston, Neoklis Polyzotis, Sudip Roy, and Steven Euijong Whang. 2016. Goods: Organizing Google's Datasets. In *SIGMOD*. 795–806.

[33] Yeye He, Kris Ganjam, and Xu Chu. 2015. Sema-join: joining semantically-related tables using big table corpora. *Proceedings of the VLDB Endowment* 8, 12 (2015), 1358–1369.

[34] Xuming Hu, Chuan Lei, Xiao Qin, Asterios Katsifodimos, Christos Faloutsos, and Huzefa Rangwala. 2025. POLYJOIN: Semantic Multi-key Joinable Table Search in Data Lakes. In *NAACL*. 384–395.

[35] Xuming Hu, Shen Wang, Xiao Qin, Chuan Lei, Zhengyuan Shen, Christos Faloutsos, Asterios Katsifodimos, George Karypis, Lijie Wen, and Philip S. Yu. 2023. Automatic Table Union Search with Tabular Representation Learning. In *ACL*. 3786–3800.

[36] Rajesh Jayaram, Laxman Dhulipala, Majid Hadian, Jason D Lee, and Vahab Mirrokni. 2024. MUVERA: Multi-Vector Retrieval via Fixed Dimensional Encoding. *NeurIPS* 37 (2024), 101042–101073.

[37] Aamod Khatiwada, Grace Fan, Roee Shraga, Zixuan Chen, Wolfgang Gatterbauer, Renée J Miller, and Mirek Riedewald. 2023. SANTOS: Relationship-based semantic table union search. *SIGMOD* 1, 1 (2023), 1–25.

[38] Aamod Khatiwada, Harsha Kokel, Ibrahim Abdelaziz, Subhajit Chaudhury, Julian Dolby, Oktie Hassanzadeh, Zhenhan Huang, Tejaswini Pedapati, Horst Samulowitz, and Kavitha Srinivas. 2025. Tabsketchfm: Sketch-based tabular representation learning for data discovery over data lakes. In *ICDE*. 1523–1536.

[39] Omar Khattab and Matei Zaharia. 2020. Colbert: Efficient and effective passage search via contextualized late interaction over bert. In *SIGIR*. 39–48.

[40] Harsha Kokel, Aamod Khatiwada, Tejaswini Pedapati, Haritha Ananthakrishnan, Oktie Hassanzadeh, Horst Samulowitz, and Kavitha Srinivas. 2025. Evaluating Joinable Column Discovery Approaches for Context-Aware Search. *arXiv preprint arXiv:2510.24599* (2025).

[41] Harsha Kokel, Aamod Khatiwada, Tejaswini Pedapati, Haritha Ananthakrishnan, Oktie Hassanzadeh, Horst Samulowitz, and Kavitha Srinivas. 2025. TOPJoin: A Context-Aware Multi-Criteria Approach for Joinable Column Search. arXiv:2507.11505 [cs.DB]

[42] Christos Koutras and Juliana Freire. 2026. StraTyper: Automated Semantic Type Discovery and Multi-Type Annotation for Dataset Collections. *arXiv preprint arXiv:2602.04004* (2026).

[43] Christos Koutras, Jiani Zhang, Xiao Qin, Chuan Lei, Vassilis N Ioannidis, Christos Faloutsos, George Karypis, and Asterios Katsifodimos. 2025. OmniMatch: Joinability discovery in data products. *Proc. VLDB Endow.* 18, 11 (2025), 4588–4601.

[44] Aditya Kusupati, Gantavya Bhatt, Aniket Rege, Matthew Wallingford, Aditya Sinha, Vivek Ramanujan, William Howard-Snyder, Kaifeng Chen, Sham Kakade, Prateek Jain, and Ali Farhadi. 2024. Matryoshka Representation Learning. arXiv:2205.13147 [cs.LG]

[45] Facebook AI Research Lab. 2015. fastText: Library for fast text representation and classification. (2015). https://fasttext.cc/

[46] Jens Lehmann, Robert Isele, Max Jakob, Anja Jentzsch, Dimitris Kontokostas, Pablo N Mendes, Sebastian Hellmann, Mohamed Morsey, Patrick Van Kleef, Sören Auer, et al. 2015. Dbpedia–a large-scale, multilingual knowledge base extracted from wikipedia. *Semantic web* 6, 2 (2015), 167–195.

[47] Oliver Lehmberg, Dominique Ritze, Robert Meusel, and Christian Bizer. 2016. A Large Public Corpus of Web Tables containing Time and Context Metadata. In *WWW*. 75–76.

[48] Peng Li, Xiang Cheng, Xu Chu, Yeye He, and Surajit Chaudhuri. 2021. Auto-fuzzyjoin: Auto-program fuzzy similarity joins without labeled examples. In *SIGMOD*. 1064–1076.

[49] Shiyuan Liu, Jianwei Wang, Xuemin Lin, Lu Qin, Wenjie Zhang, and Ying Zhang. 2026. HyperJoin: LLM-augmented Hypergraph Link Prediction for Joinable Table Discovery. *arXiv preprint arXiv:2601.01015* (2026).

[50] Marc Maynou, Sergi Nadal, Raquel Panadero, Javier Flores, Oscar Romero, and Anna Queralt. 2026. FREYJA: Efficient join discovery in data lakes. *IEEE Trans. Knowl. Data Eng.* 38 (2026), 1–12.

[51] Leland McInnes, John Healy, Nathaniel Saul, and Lukas Großberger. 2018. UMAP: Uniform Manifold Approximation and Projection. *Journal of Open Source Software*

3, 29 (2018), 861. https://doi.org/10.21105/joss.00861
[52] Renée J. Miller. 2018. Open Data Integration. *Proc. VLDB Endow.* 11, 12 (2018), 2130–2139.
[53] Renée J. Miller, Fatemeh Nargesian, Erkang Zhu, Christina Christodoulakis, Ken Q. Pu, and Periklis Andritsos. 2018. Making Open Data Transparent: Data Discovery on Open Data. *IEEE Data Eng. Bull.* 41, 2 (2018), 59–70.
[54] Pranay Mundra, Jianhao Zhang, Fatemeh Nargesian, and Nikolaus Augsten. 2023. Koios: Top-k semantic overlap set search. In *ICDE*. 1531–1543.
[55] Fatemeh Nargesian, Ken Q. Pu, Bahar Ghadiri Bashardoost, Erkang Zhu, and Renée J. Miller. 2023. Data Lake Organization. *IEEE Trans. Knowl. Data Eng.* 35, 1 (2023), 237–250.
[56] Fatemeh Nargesian, Ken Q. Pu, Erkang Zhu, Bahar Ghadiri Bashardoost, and Renée J. Miller. 2020. Organizing Data Lakes for Navigation. In *SIGMOD*. 1939–1950.
[57] Fatemeh Nargesian, Erkang Zhu, Renée J. Miller, Ken Q. Pu, and Patricia C. Arocena. 2019. Data Lake Management: Challenges and Opportunities. *Proc. VLDB Endow.* 12, 12 (2019), 1986–1989.
[58] Fatemeh Nargesian, Erkang Zhu, Ken Q. Pu, and Renée J. Miller. 2018. Table Union Search on Open Data. *Proc. VLDB Endow.* 11, 7 (2018), 813–825.
[59] Arash Dargahi Nobari and Davood Rafiei. 2022. Efficiently transforming tables for joinability. In *ICDE*. 1649–1661.
[60] OpenAI. 2026. GPT-5.5. https://developers.openai.com/api/docs/models/gpt-5.5.
[61] Paul Ouellette, Aidan Sciortino, Fatemeh Nargesian, Bahar Ghadiri Bashardoost, Erkang Zhu, Ken Pu, and Renée J. Miller. 2021. RONIN: Data Lake Exploration. *Proc. VLDB Endow.* 14, 12 (2021), 2863–2866.
[62] Fabian Pedregosa, Gaël Varoquaux, Alexandre Gramfort, Vincent Michel, Bertrand Thirion, Olivier Grisel, Mathieu Blondel, Andreas Müller, Joel Nothman, Gilles Louppe, Peter Prettenhofer, Ron Weiss, Vincent Dubourg, Jake Vanderplas, Alexandre Passos, David Cournapeau, Matthieu Brucher, Matthieu Perrot, and Édouard Duchesnay. 2018. Scikit-learn: Machine Learning in Python. arXiv:1201.0490 [cs.LG]
[63] D. Ritze, O. Lehmberg, R. Meusel, C. Bizer, and S. Zope. 2015. WDC Web Table Corpus. http://webdatacommons.org/webtables/2015/downloadInstructions.html
[64] Aécio Santos, Aline Bessa, Fernando Chirigati, Christopher Musco, and Juliana Freire. 2021. Correlation sketches for approximate join-correlation queries. In *SIGMOD*. 1531–1544.
[65] Aécio S. R. Santos, Aline Bessa, Christopher Musco, and Juliana Freire. 2022. A Sketch-based Index for Correlated Dataset Search. In *ICDE*. 2928–2941.
[66] Aécio S. R. Santos, Flip Korn, and Juliana Freire. 2024. Efficiently Estimating Mutual Information Between Attributes Across Tables. In *ICDE*. 193–206.
[67] Anish Das Sarma, Lujun Fang, Nitin Gupta, Alon Y. Halevy, Hongrae Lee, Fei Wu, Reynold Xin, and Cong Yu. 2012. Finding related tables. In *SIGMOD*. 817–828.
[68] Kaitao Song, Xu Tan, Tao Qin, Jianfeng Lu, and Tie-Yan Liu. 2020. Mpnet: Masked and permuted pre-training for language understanding. *NeurIPS* 33 (2020), 16857–16867.
[69] Henrique Schechter Vera, Sahil Dua, Biao Zhang, Daniel Salz, Ryan Mullins, Sindhu Raghuram Panyam, Sara Smoot, Iftekhar Naim, Joe Zou, Feiyang Chen, et al. 2025. Embeddinggemma: Powerful and lightweight text representations. *arXiv preprint arXiv:2509.20354* (2025).
[70] WDC. [n.d.]. http://webdatacommons.org/webtables/goldstandard.html, last accessed on Feb 15, 2026.
[71] Shitao Xiao, Zheng Liu, Peitian Zhang, and Niklas Muennighoff. 2023. C-Pack: Packaged Resources To Advance General Chinese Embedding. arXiv:2309.07597 [cs.CL]
[72] Haoxiang Zhang, Yurong Liu, Aécio Santos, Wei-Lun (Allen) Hung, and Juliana Freire. 2026. AutoDDG: Automated Dataset Description Generation using Large Language Models. *Proc. ACM Manag. Data* 4, 1, Article 12 (April 2026), 27 pages. https://doi.org/10.1145/3786626
[73] Zixuan Zhao and Raul Castro Fernandez. 2022. Leva: Boosting Machine Learning Performance with Relational Embedding Data Augmentation. In *SIGMOD*. 1504–1517.
[74] Erkang Zhu, Dong Deng, Fatemeh Nargesian, and Renée J. Miller. 2019. JOSIE: Overlap Set Similarity Search for Finding Joinable Tables in Data Lakes. In *SIGMOD*. 847–864.
[75] Erkang Zhu, Yeye He, and Surajit Chaudhuri. 2017. Auto-join: Joining tables by leveraging transformations. *Proc. VLDB Endow.* 10, 10 (2017), 1034–1045.
[76] Erkang Zhu, Fatemeh Nargesian, Ken Q. Pu, and Renée J. Miller. 2016. LSH Ensemble: Internet-Scale Domain Search. *Proc. VLDB Endow.* 9, 12 (2016), 1185–1196.